\documentclass[%
 reprint,
 amsmath,amssymb,
 aps,superscriptaddress
]{revtex4-2}

\newcommand{\beginsupplement}{%
        \setcounter{table}{0}
        \renewcommand{\thetable}{S\arabic{table}}%
        \setcounter{figure}{0}
        \renewcommand{\thefigure}{S\arabic{figure}}%
     }
\usepackage{graphicx}
\usepackage{dcolumn}
\usepackage[colorlinks=true,linkcolor=blue,urlcolor=blue,citecolor=blue]{hyperref}
\usepackage{bm}
\usepackage{float}
\usepackage{xcolor}
\usepackage{graphicx}
\usepackage{float}
\begin{document}

\preprint{APS/123-QED}

\title{Non-monotonic skewness of currents in non-equilibrium steady states
}

\author{Sreekanth K Manikandan$^\dagger$}\email{sreekanth.km@fysik.su.se}
\affiliation{NORDITA, KTH Royal institute of technology and Stockholm university, Stockholm.}
\author{Biswajit Das}\thanks{These authors contributed equally}
\author{Avijit Kundu}\thanks{These authors contributed equally}
\author{Raunak Dey}
\affiliation{Department of Physical Sciences, Indian Institute of Science Education and Research Kolkata, Mohanpur Campus, Mohanpur, West Bengal 741246, India}
\author{Ayan Banerjee}\email{ayan@iiserkol.ac.in}
\affiliation{Department of Physical Sciences, Indian Institute of Science Education and Research Kolkata, Mohanpur Campus, Mohanpur, West Bengal 741246, India}%
\author{Supriya Krishnamurthy}\email{supriya@fysik.su.se}
\affiliation{Department of Physics, Stockholm university, Stockholm.}

\begin{abstract}
 Measurements of any property of a microscopic system are bound to  show significant deviations from the average, due to thermal fluctuations. For time-integrated currents such as heat, work or entropy production in a steady state,  it is in fact known that there will be long stretches of fluctuations both above as well as  below the average, occurring equally likely at large times. In this paper we show that for any finite-time measurement in a non-equilibrium steady state  - rather counter-intuitively -  fluctuations below the average are more probable. This discrepancy is higher when the system is further away from equilibrium.  For overdamped diffusive processes, there is even an optimal time when time-integrated current fluctuations mostly lie below the average.  We demonstrate that these effects result from the non-monotonic skewness of current fluctuations and provide evidence that they are easily observable in experiments. We also discuss
their extensions to discrete space Markov jump processes
and implications to biological and synthetic microscopic
engines.
\end{abstract}

\maketitle
 In microscopic non-equilibrium systems, individual measurements of heat, work or entropy production can significantly fluctuate about the average values \cite{bustamante2005nonequilibrium}. The nature of these fluctuations are constrained within the framework of stochastic thermodynamics by some universal results \cite{seifert2012stochastic}. The most celebrated ones are  the fluctuation theorems (see review \cite{jarzynski2011equalities} and references therein) which constrain the probability distributions of thermodynamic quantities such as heat, work and entropy production. Its applications range from estimating Free-energy differences in single molecule experiments \cite{hummer2001free,liphardt2002equilibrium} to determining the nature of efficiency fluctuations in microscopic engines \citep{verley2014unlikely,verley2014universal,manikandan:EF}. 
Another class of results provide bounds on the fluctuations of \textit{currents} in non-equilibrium steady states in terms of the steady state entropy production rate $\sigma = \langle \Delta S_{tot}\rangle/t$ \cite{dba}. 
For any current $J$ in a stationary state of a continuous time, Markov process, it can be shown that the scaled cumulant generating function $\phi_J^\sigma (\lambda,t)\equiv\frac{1}{t}\log \langle e^{-\lambda \sigma t \frac{J}{\langle J \rangle}}\rangle_t$ is bounded from below by a parabola \cite{dba,univc,pft}, 
\begin{align}
\label{eq:bound}
    \phi(\lambda,t)\geq -\sigma \lambda(1-\lambda).
\end{align}
Terminating the { expansion of $\phi$ } to the second order in $\lambda$, leads to the thermodynamic uncertainty relations \cite{TURprl,TURnatphys,TURannrev} which are trade-off relations connecting the precision of arbitrary current measurements to the entropy production rate. 

In the $t\rightarrow \infty$ limit, the LHS of Eq.\ \eqref{eq:bound} converges to a time independent function \cite{dba} referred to as the large deviation rate function \cite{ellis2006entropy,touchette2009large,lebowitz1999gallavotti,BernardD,chetrite2015nonequilibrium}, knowing which helps fully characterize the fluctuations in the long-time limit.
In general, such long-time results  can be obtained within the mathematical framework of large deviation theory and many such results have been obtained
for the statistics of the fluctuations of entropy production \cite{mehl2008large,speck2012large,kundu2011large,sabhapandit2012heat,verley2014work,NewRef}, efficiency distributions \cite{verley2014unlikely,verley2014universal,manikandan:EF}, first passage problems \cite{saito2016waiting,simplebounds,singh2019generalised} and current fluctuations in general \cite{additivity,derrida2007non}. An interesting addition to this class of results was obtained in Ref.\ \cite{barato2018arcsine}, where it was shown that the fraction of time that a current spends above its average value follows the arcsine law in the long time limit \cite{levy1940certains}. As a consequence,
stochastic currents with long streaks above or below their average value are much more and equally likely than those that spend similar fractions of time above and below their average. 

The other extreme of very short-time fluctuations of currents, is also surprisingly non-trivial. It has been shown that Eq.\ \eqref{eq:bound} saturates for $J=\Delta S_{tot}$ in the limit $t\rightarrow 0$ for overdamped diffusive processes \cite {manikandan2020inferring,shun,tan,manikandan2021quantitative}. As a consequence, $\sigma$ can be exactly inferred for such systems by studying the mean and variance of current fluctuations at short times \cite {manikandan2020inferring,shun,tan,manikandan2021quantitative}, even for non-stationary systems \cite{otsubo2020estimating}.   The saturation of the bound also implies that the fluctuations of $\Delta S_{tot}$ are Gaussian in these systems in the short-time limit even when arbitrarily far from equilibrium \cite{manikandan2020inferring}. In fact, as was recently pointed out, the Gaussianity in the $t\rightarrow 0$ limit holds for any arbitrary current in overdamped diffusive processes \cite{otsubo2020estimating}. As we show below, this short-time behaviour combined with the large-deviation results mentioned above hold important clues for interesting finite-time fluctuation properties.

Finite-time fluctuations are clearly of interest since this is most often what is observed in experiments.
However, when neither the $t\rightarrow \infty $ nor the $t\rightarrow 0$ limit can be taken, generic features of such fluctuations are harder to identify  because of the prevalence of transient effects and time-correlations. If the $t\rightarrow \infty $ limit can be thought  in terms of applying a  thermodynamic limit \citep{jack2020ergodicity}, then finite-time fluctuations include effects which vanish in the thermodynamic limit, making them harder to access. Some important general results in this category 
include the integral fluctuation theorem for stochastic entropy production \citep{seifertep}, statistical properties of entropy production derivable from the fluctuation theorem \cite{merhav2010statistical},  the finite-time versions of the thermodynamic uncertainty relations \cite{ftg,pft,dechant2018current,flcunc}, universal results known for the statistics of infima, stopping times, and first-passage probabilities of entropy production \cite{neri}, a generic equation describing the  time evolution of the stochastic entropy production \cite{generic}, statistics of the time of the maximum of a one dimensional stationary process \cite{mori2021distribution}, and bounds on first passage times of current fluctuations \cite{firstpassage}. 

In this paper, we unravel a previously unnoticed property of current fluctuations at finite and short-times.
We demonstrate that the \textit{skewness} of current fluctuations is positive, and non-monotonic in time
and argue that this behaviour is generic for non-equilibrium steady states generated by any overdamped diffusive process. 
As a consequence, 
 we find that at all finite times, current fluctuations below the long-time average are more probable than those above. For a single realization of the process, interestingly, this implies that a below average outcomes will be typical.  Moreover, due to the non-monotonicity, there is an optimal time when this discrepancy is the highest. 
We show that these features of current fluctuations are easily visible in non-trivial models studied numerically and experimentally. We also discuss their extensions to discrete space Markov jump processes and implications to biological and synthetic microscopic engines. In all cases, in the limit of large $t$, we recover results consistent with Ref.\ \cite{barato2018arcsine}. 

The central results we present in this manuscript apply to non-equilibrium systems in a stationary state. We first consider generic overdamped diffusive processes of the form, 
\begin{align}
    \dot{{\bm x}}(t)={\bm A}({\bm x}(t))+{\bm B}({\bm x}(t), t)\cdot {\bm \eta}(t),\label{eq: Langevin}
\end{align}
where ${\bm A}({\bm x})$ is the drift vector, and ${\bm B}({\bm x}, t)$ is a $d \times d$ matrix, and ${\bm \eta}(t)$ represents a Gaussian white noise satisfying $\langle \eta_i(t) \eta_j(t')\rangle = \delta_{ij}\delta(t-t')$. 
Consider a current $J$ in the stationary state of this system defined as $J =\int_{\textbf{x}(0)}^{\textbf{x}(t)} \textbf{d}(\textbf{x})\circ d\textbf{x}$, where ${\bf d}({\bf x})$ is any arbitrary function of ${\bf x}$, and $\circ$ corresponds to the Stratanovich product. First we look at the implications of Eq.\ \eqref{eq:bound} for any such current. Without loss of generality, we consider currents for which $\langle J \rangle \geq 0$. Let $\langle\left[J(t)\right]^k\rangle$ be the $k$-th  cumulant of $J$ \footnote{Note Marcinkiewicz theorem \cite{marcinkiewicz1939propriete}, which implies that if a distribution is not Gaussian, then it will have further non-zero cumulants.}. Expanding the inequality in Eq.\ \eqref{eq:bound} in powers of $(-\lambda)^i$, it can be shown that $\langle\left[J(t)\right]^k\rangle \geq 0 $ for $k \geq 2$ for any $t$. For $1 < k \leq 3$, the cumulants coincide with the $k-$th central moment of  $J$. 
Here $\langle \cdot \rangle$ corresponds to an ensemble average over steady state trajectories of length $t$.  An important standardized moment which can then be constructed is the skewness, $S = \dfrac{\langle\left[J(t)\right]^3\rangle}{\langle\left[J(t)\right]^2\rangle^{3/2}}$. Skewness quantifies the asymmetries of the fluctuations about the average value. Here we focus on the properties of the skewness as a function of $t$. Using the results in Ref. \cite{otsubo2020estimating} which proved the Gaussianity of general current fluctuations in the $t\rightarrow 0$ limit (see Eq. S13 - S15 in the Supplementary Note 1 of \cite{otsubo2020estimating}, where it is shown that $\langle J(t) \rangle$, and $\langle J(t)^2 \rangle$ are $\propto t$ for small $t$, but $\langle J(t)^3 \rangle \propto t^2$ for small $t$. For $J=\Delta S_{tot}$, this behaviour was shown already in \cite{manikandan2020inferring}), we obtain that $S\propto t^{1/2}$ for small $t$. 
Further, the existence of the large deviation function $ \Phi (\lambda)=\lim_{t\rightarrow \infty} \phi(\lambda,t)$ ensures that all the cumulants scale linearly in time as $t\rightarrow \infty$. As a result, $S\propto t^{-1/2}$ for large $t$.   Combining these two limiting behaviours with the positivity of the cumulants, we obtain that $S$ is a  positive, non-monotonic function of time that vanishes both in the $t \rightarrow \infty $ limit as well as the $t \rightarrow 0$ limit. As a consequence, there will also be a special time, where the skewness attains a maximum value. This is the first central observation we make in this paper.

From the generality of the above arguments, we expect this behaviour to be generic for any  non-equilibrium steady state current if the short-time and long-time behaviour are as detailed above. But to be more concrete, we now take the example of the 
non-equilibrium steady state of a colloidal system \cite{gomez:ssw,Pal:2013wfb,verley2014work,Manikandan:2017awd,manikandan2018exact,manikandan2021quantitative}. The model consists of a single colloidal particle in a harmonic trap with stiffness $\kappa$, whose mean position is modulated according to the Ornstein-Uhlenbeck process. The dynamics of the system with position variable $x(t)$, and the trap center $x_0(t) \equiv \lambda(t)$ can be described using a system of overdamped Langevin equations as:
\begin{align}
\label{eq:lgv}
\begin{split}
    \dot{x}(t) &= -\dfrac{x(t)-\lambda(t)}{\tau}+\sqrt{2D}\zeta(t)\\
    \dot{\lambda}(t) &= -\dfrac{\lambda(t)}{\tau_0}+\sqrt{2A}\xi(t)
    \end{split}
\end{align}
Here $D$ is the diffusion constant at room temperature ($T$), $\tau=\gamma/\kappa$ is the relaxation time of the harmonic trap,  $\kappa$ is the trap stiffness and $\gamma$ is the drag coefficient related to $D$ by the Stokes-Einstein relation as $D\gamma=k_BT$. Similarly, $\tau_0$ is the relaxation time of the OU process and $A$ corresponds to its strength. The noise terms $\zeta(t)$ and $\xi(t)$ are Gaussian white noises obeying $\langle\xi(t)\rangle = 0$, $\langle\zeta(t)\rangle = 0$,  $\langle\xi(t)\xi(t')\rangle=\delta(t-t^\prime)$, $\langle\zeta(t)\zeta(t')\rangle=\delta(t-t^\prime)$ and $\langle\xi(t)\zeta(t')\rangle=0$. 
 In Fig.~1, we plot the skewness $S$ of several arbitrary currents as a function of time, using numerical data. Currents are constructed using $J =\int_{\textbf{x}(0)}^{\textbf{x}(t)} \textbf{d}(\textbf{x})\circ d\textbf{x}$, where $\bf{x}$=$[x;\lambda]^T$, and $
    {\bf d}({\bf x}) = \left[\begin{array}{c}
         c_1x+c_2 \lambda;\;  
         c_3x+c_4\lambda 
    \end{array} \right]^T$. Here $\lbrace c_i \rbrace \in \mathbb{R}$ are constants which can be varied to construct different currents. Without loss of generality, we consider currents with $\langle J \rangle \geq 0$. We find that the skewness of arbitrary currents is a positive, bounded, and non-monotonic function of $t$ which vanishes both in the short -time limit as $t^{1/2}$ and in the large-time limit as $t^{-1/2}$. 

\begin{figure}
    \centering
       \includegraphics[scale=0.54]{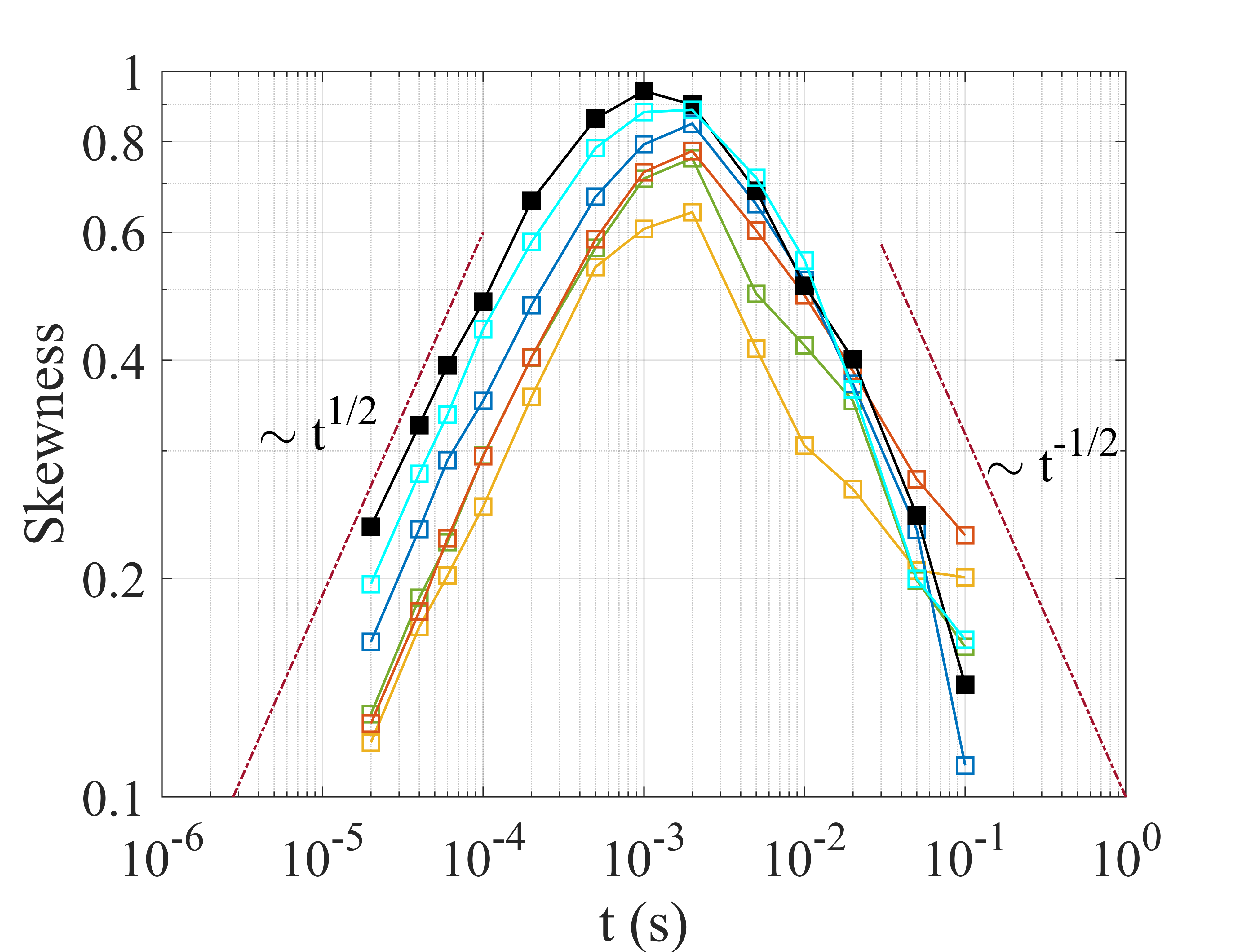}
    \caption{ The skewness $S$  of arbitrary currents (shown in colors) in non-equilibrium steady state of the stochastic sliding parabola model as a function of $t$. The vanishing of $S(t)$ in the $t\rightarrow 0$ limits shows the emergence of Gaussian fluctuations in the short-time limit. The solid black line corresponds to $J=\Delta S_{tot}$. The red dashed lines corresponds to power-law fits at the short and large $t$ limits. }
    \label{fig:1}
\end{figure}

\begin{figure*}
    \centering
    \includegraphics[width=18cm]{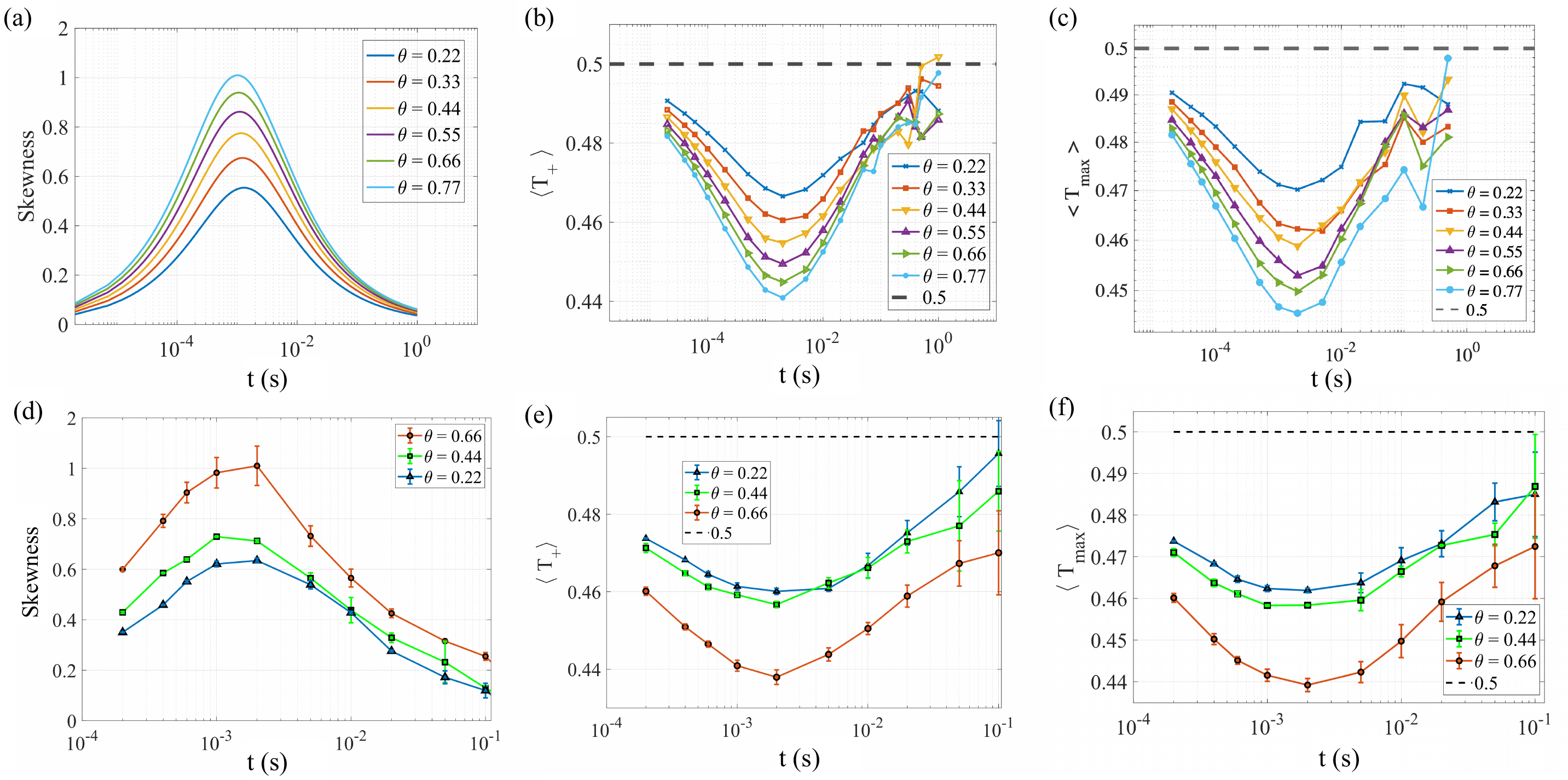}
    \caption{a) Skewness of $\Delta S_{tot}$ as a function of $t$ computed analytically for the model in Eq.\ \eqref{eq: Langevin}. b) $\langle T_+ \rangle$ and c) $\langle T_{max} \rangle$  as a function of $t$ for different values of $\theta$, obtained from numerical simulation of the system. In the bottom panel, (d) to (f), we show the same results obtained from the experimental realization of the same system.}
    \label{fig:2}
\end{figure*}

Next, we investigate these properties for a particular choice of the current, which is $J = \Delta S_{tot}$ \footnote{See Eq.\ (14b) in Ref.\ \cite{manikandan2021quantitative} for the choice of $c_i$'s that define $J = \Delta S_{tot}$}. 
The corresponding entropy production rate is given by  $\sigma = \frac{\delta^2\theta}{(\delta+1)\tau_0}$,
where $\delta=\frac{\tau_0}{\tau}$ and $\theta =\frac{A}{D}$ \cite{Pal:2013wfb,verley2014work,manikandan2018exact}.
In Fig.\ \ref{fig:2}a, we plot the analytically computed skewness of $\Delta S_{tot}(t)$, for different values of $\theta$ \footnote{The exact analytic calculation involves the computation of the finite-time moment generating function $G(\lambda) = \langle e^{-\lambda \Delta S_{tot}}\rangle_t$ using a path integral technique which was developed in \cite{Manikandan:2017awd} and used in  \cite{manikandan2018exact}. For completeness, we reproduce the results (from \cite{manikandan2018exact})in the supplemental material.}.
As expected, we find that $S(t) \geq 0 $ and is non-monotonic in time, featuring a maximum at an intermediate time $t\equiv \tau_E$. 
We also find that the skewness increases with $\theta$, which also increases the entropy production rate of the system. 

Now we look at the measurable consequences of this non-monotonicity. 
We first consider the mean fraction of the time when the measured entropy stays above the average value, denoted by $\langle T_+ \rangle = \frac{1}{t}\int_{0}^t \Theta (J(s)-\langle J(s)\rangle) \;ds$,
where $\Theta(x)$ is the Heaviside function.
 If the fluctuations of $\Delta S_{tot}(t)$ were  symmetric about the average, then it would have implied, $\langle T_+ \rangle = \frac{1}{2}$. Indeed, this is known to be the case in the $t\rightarrow \infty$ limit \cite{barato2018arcsine}. In Fig.\ \ref{fig:2}b, we plot $\langle T_+ \rangle$, obtained from the numerical data as a function of $t$. 
 We find that,   $\langle T_+ \rangle < \frac{1}{2}$ for all $t$ and tends to $\frac{1}{2}$ in the limits $t\rightarrow 0$ and $t\rightarrow \infty$. We also find that  $\langle T_+ \rangle$ is non-monotonic in time, and attains a minimum close to when $S(t)$ is the highest. 

 Interestingly, the behaviour of $\langle T_+ \rangle $ implies that  it is more likely that a current measured for a finite $t$ duration stays below the average value for most part of the measurement. This discrepancy is the highest when the skewness peaks.  We also find that $\langle T_+ \rangle$ is monotonically decreasing as a function of $\theta$  for all $t$. Hence,  the further away the system is from equilibrium, the more likely that arbitrary currents or entropy production measured along a single trajectory stay below the average value for most of the time.
 
 Next we consider $\langle T_{max}(t) \rangle$, where  $T_{max}(t)=\frac{t_{sup}}{t}$, where $t_{sup}$ is the time of global maximum of $\Delta S_{tot}(t)-\sigma t$. The results are shown in Fig.\ \ref{fig:2}c. We find that $\langle T_{max} \rangle$ has a similar time dependence as  $\langle T_+ \rangle$ and stays  below $\frac{1}{2}$ for all $t$ and tends to $\frac{1}{2}$ in the limits $t\rightarrow 0$ and $t\rightarrow \infty$.   $\langle T_{max} (t)\rangle$ is also found to be non-monotonic in time featuring a minimum at $t\sim \tau_E$. This means, for currents measured for a finite time, it is more likely that it's maximum deviation above the mean will be found  before the half-length of the measurement. Again, the discrepancy will be the highest when $t\sim \tau_E$. 

As seen in Fig.~1, the time at which the skewness has the highest value varies from current to current. In particular, for $J=\Delta S_{tot}$, from the analytical solutions, we find that $\tau_E$ monotonically decreases with increasing $\theta$ (See the supplementary information where the dependence of $\tau_E$ on $\theta$ and $\tau_0$ is analytically obtained). As a result, in order to capture the non monotonic nature of $S(t)$, the further a system is away from equilibrium, the finer the resolution needs to be. We demonstrate below that this is, however, still very detectable in experimental data.


\textbf{Experiments:} 
The model in Eq.\ \eqref{eq:lgv} was first realized in an Optical Tweezers setup in Ref.\ \cite{gomez:ssw} and was also studied recently in Ref.\ \cite{manikandan2021quantitative}. To realize this system, we trap a 3 $\mu$m polystyrene particle in an aqueous solution in a harmonic potential well given by $U(x(t),\lambda(t))=k[x(t)-\lambda(t)]^2/2$,  where the trap stiffness is $k=19.7\pm0.1$ pN/$\mu$m. Here $\lambda(t)$ is the time dependent mean position of the trap which is modulated using an acousto-optic modulator according to an Ornstein-Uhlenbeck process (see Eq~\ref{eq:lgv}) with $\tau_0=2.5$ ms, and $A= \left[0.1,0.2,0.3\right]\times(0.6\times10^{-6})^2\ m^2/s$. We sample the one-directional trajectory of the probe at a spatio-temporal resolution of $\sim 1$ nm - 10 kHz for 100 second. We then use the autocorrelation of the time series of a trapped particle and the noise to calibrate the fluctuation of the probe from volts 
to nm \cite{bera2017scirep}.  We plot the experimental results in Fig.\ \ref{fig:2}d, e and f. We find that the numerical results in Fig.\ \ref{fig:2} (a)-(c) are reproduced by our experiments and that the non-monotonicities in $S(t)$, $\langle T_+ \rangle$ and $\langle T_{max} \rangle$ are clearly visible \footnote{Parameters used in the plots are given in section IV of supplemental material \cite{supp}}.

\begin{figure*}
    \centering
    \includegraphics[width=\textwidth]{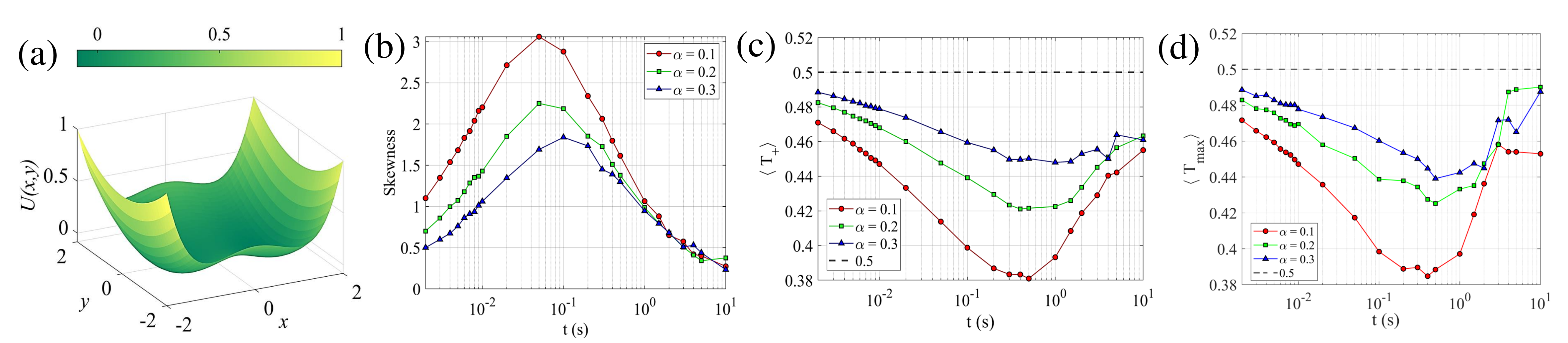}
    \caption{   a) Double-well potential with $b$ = 1 and $k$ = 2  (See the details in \cite{das2022inferring}). (b) Skewness ,(c) $\langle T_+ \rangle$ and d) $\langle T_{max} \rangle$ of the entropy currents as a function of $t$, obtained numerically for the gyrator model with double-well confining potential for different values of the ratio of the temperatures ($\alpha = \frac{T_2}{T_1}$) along the two orthogonal directions of the gyrator system. See the supplementary information \cite{supp} for relevant details of the model.}
    \label{fig:plot_nonlinear}
\end{figure*}
So far, we have considered a linear Langevin model. However, our results are generic and are expected to hold for any overdamped diffusive process. To substantiate this, we numerically explore the non-monotonic nature of fluctuations of entropy currents of a system with non-linearities. For this, we consider an anharmonic Brownian gyrator with a double-well confining potential (studied recently in \cite{das2022inferring}; see the supplemental material for details and for the additional example of a gyrator with a quartic confining potential). We vary the non-equilibrium conditions by changing  the ratio of the temperatures ($\alpha = \frac{T_2}{T_1}$) along the two orthogonal directions.
In Fig.~\ref{fig:plot_nonlinear}, we show that all the features that we showed in Fig.\ \ref{fig:2} are also present in this non-linear model.

\begin{figure*}
    \centering
    \includegraphics[width=18cm]{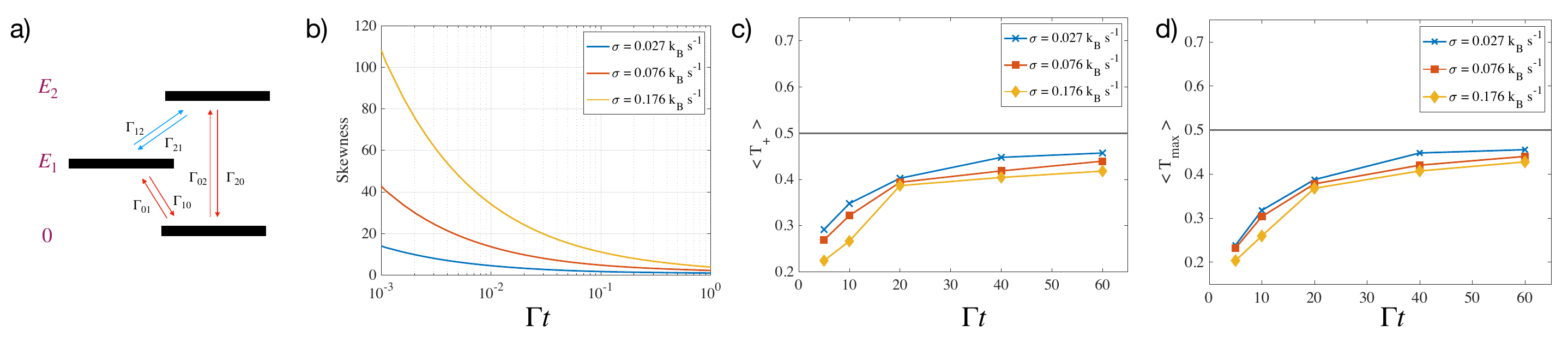}
    \caption{a) Three level system. b) Skewness of $\Delta S_{tot}$, c) $\langle T_+ \rangle$ and d) $\langle T_{max} \rangle$ as a function of $t$ computed analytically for the three level system. See the supplemental material  \cite{supp} for details and the choice of the parameters.}
    \label{fig:5}
\end{figure*}

Apart from the generic bound in Eq.\ \eqref{eq:bound} which holds for any continuous-time Markov process, a crucial ingredient in our results is the emergence of Gaussian fluctuations in the short-time limit of overdamped diffusive systems, recently proved in \cite{otsubo2020estimating}. For discrete-space systems, which evolve according to a continuous-time Markov process, current fluctuations are not necessarily Gaussian at $t\rightarrow 0$; in fact it can be shown that all the moments scale $\propto t$ at short times  \cite{supp}. Thus, the skewness will scale $\propto t^{-1/2}$,  diverging as $t\rightarrow 0$, and will not necessarily be non-monotonic in time. As a consequence, shorter current measurements in such systems are more likely to lie below the average.
In all the cases these effects increase when the system is further away from equilibrium.
In Fig.\ \ref{fig:5}, we demonstrate this for a three-level system, coupled to two thermal reservoirs at unequal temperatures \cite{PhysRevLett.2.262,PhysRevLett.119.050602,PhysRevLett.122.110601}. Variants of this system have recently appeared in a number of different contexts \cite{PhysRevLett.128.140602,singh2020optimal,PhysRevResearch.3.043108}, an important example being classical/ quantum  clocks \cite{milburn2020thermodynamics}. It is known that clocks need to run for a long time to give reliable estimates \footnote{see discussion around Eq. (92) in \cite{milburn2020thermodynamics}} which is also the limit in which the skewness vanishes.

In summary, we have unravelled and quantitatively characterized the universal properties of  skewness of current fluctuations, and their measurable consequences in non-equilibrium steady states. 
The results imply that, rather counter intuitively, it is more probable that currents such as entropy production will mostly lie below average in a single realization of finite-time duration. Our results can be potentially verified in molecular motors such as kinesin \cite{vale1997design} by looking at the statistics of it's steps \cite{schnitzer1997kinesin} or energy dissipation \cite{ariga2018nonequilibrium}.  Our results also show that the nature of the fluctuations of currents, and thus the most probable outcome, crucially depend on the time duration of the measurement. It would be of substantial interest to investigate whether biological motors have optimized their timescales considering such constraints (through molecular evolution).
For artificial microscopic engines such as the ones studied in Refs.~\cite{blickle2012realization,martinez2016brownian,filliger2007brownian,chiang2017electrical,chang2021autonomous,argun2017experimental}, this implies that it might be possible to choose cycling times to optimize fluctuations, and to get a reliable performance in a limited number of runs. We plan to attempt answering these questions in our future research.

\vspace{2mm}
\section*{Author contribution Statement}
SKM and SK developed the theory. AK and BD designed and performed the experiments in AB's Lab. SKM, RD and BD analyzed the numerical and experimental data. 
\section*{Acknowledgements}
SK acknowledges the support of the Swedish Research  Council  through the grant 2021-05070. BD is thankful to Ministry Of Education of Government of India for the financial support through the Prime Minister's Research Fellowship (PMRF) grant. AK acknowledges the DST, Govt. of India for INSPIRE Fellowship.
\section*{Data and Code Availability}
The raw data files and codes are available openly at \href{https://doi.org/10.6084/m9.figshare.17703269.v2}{https://doi.org/10.6084/m9.figshare.17703269.v2} 


\begin{thebibliography}{81}%
\makeatletter
\providecommand \@ifxundefined [1]{%
 \@ifx{#1\undefined}
}%
\providecommand \@ifnum [1]{%
 \ifnum #1\expandafter \@firstoftwo
 \else \expandafter \@secondoftwo
 \fi
}%
\providecommand \@ifx [1]{%
 \ifx #1\expandafter \@firstoftwo
 \else \expandafter \@secondoftwo
 \fi
}%
\providecommand \natexlab [1]{#1}%
\providecommand \enquote  [1]{``#1''}%
\providecommand \bibnamefont  [1]{#1}%
\providecommand \bibfnamefont [1]{#1}%
\providecommand \citenamefont [1]{#1}%
\providecommand \href@noop [0]{\@secondoftwo}%
\providecommand \href [0]{\begingroup \@sanitize@url \@href}%
\providecommand \@href[1]{\@@startlink{#1}\@@href}%
\providecommand \@@href[1]{\endgroup#1\@@endlink}%
\providecommand \@sanitize@url [0]{\catcode `\\12\catcode `\$12\catcode
  `\&12\catcode `\#12\catcode `\^12\catcode `\_12\catcode `\%12\relax}%
\providecommand \@@startlink[1]{}%
\providecommand \@@endlink[0]{}%
\providecommand \url  [0]{\begingroup\@sanitize@url \@url }%
\providecommand \@url [1]{\endgroup\@href {#1}{\urlprefix }}%
\providecommand \urlprefix  [0]{URL }%
\providecommand \Eprint [0]{\href }%
\providecommand \doibase [0]{https://doi.org/}%
\providecommand \selectlanguage [0]{\@gobble}%
\providecommand \bibinfo  [0]{\@secondoftwo}%
\providecommand \bibfield  [0]{\@secondoftwo}%
\providecommand \translation [1]{[#1]}%
\providecommand \BibitemOpen [0]{}%
\providecommand \bibitemStop [0]{}%
\providecommand \bibitemNoStop [0]{.\EOS\space}%
\providecommand \EOS [0]{\spacefactor3000\relax}%
\providecommand \BibitemShut  [1]{\csname bibitem#1\endcsname}%
\let\auto@bib@innerbib\@empty
\bibitem [{\citenamefont {Bustamante}\ \emph {et~al.}(2005)\citenamefont
  {Bustamante}, \citenamefont {Liphardt},\ and\ \citenamefont
  {Ritort}}]{bustamante2005nonequilibrium}%
  \BibitemOpen
  \bibfield  {author} {\bibinfo {author} {\bibfnamefont {C.}~\bibnamefont
  {Bustamante}}, \bibinfo {author} {\bibfnamefont {J.}~\bibnamefont
  {Liphardt}},\ and\ \bibinfo {author} {\bibfnamefont {F.}~\bibnamefont
  {Ritort}},\ }\bibfield  {title} {\bibinfo {title} {The nonequilibrium
  thermodynamics of small systems},\ }\href@noop {} {\bibfield  {journal}
  {\bibinfo  {journal} {arXiv preprint cond-mat/0511629}\ } (\bibinfo {year}
  {2005})}\BibitemShut {NoStop}%
\bibitem [{\citenamefont {Seifert}(2012)}]{seifert2012stochastic}%
  \BibitemOpen
  \bibfield  {author} {\bibinfo {author} {\bibfnamefont {U.}~\bibnamefont
  {Seifert}},\ }\bibfield  {title} {\bibinfo {title} {Stochastic
  thermodynamics, fluctuation theorems and molecular machines},\ }\href@noop {}
  {\bibfield  {journal} {\bibinfo  {journal} {Reports on progress in physics}\
  }\textbf {\bibinfo {volume} {75}},\ \bibinfo {pages} {126001} (\bibinfo
  {year} {2012})}\BibitemShut {NoStop}%
\bibitem [{\citenamefont {Jarzynski}(2011)}]{jarzynski2011equalities}%
  \BibitemOpen
  \bibfield  {author} {\bibinfo {author} {\bibfnamefont {C.}~\bibnamefont
  {Jarzynski}},\ }\bibfield  {title} {\bibinfo {title} {Equalities and
  inequalities: Irreversibility and the second law of thermodynamics at the
  nanoscale},\ }\href@noop {} {\bibfield  {journal} {\bibinfo  {journal} {Annu.
  Rev. Condens. Matter Phys.}\ }\textbf {\bibinfo {volume} {2}},\ \bibinfo
  {pages} {329} (\bibinfo {year} {2011})}\BibitemShut {NoStop}%
\bibitem [{\citenamefont {Hummer}\ and\ \citenamefont
  {Szabo}(2001)}]{hummer2001free}%
  \BibitemOpen
  \bibfield  {author} {\bibinfo {author} {\bibfnamefont {G.}~\bibnamefont
  {Hummer}}\ and\ \bibinfo {author} {\bibfnamefont {A.}~\bibnamefont {Szabo}},\
  }\bibfield  {title} {\bibinfo {title} {Free energy reconstruction from
  nonequilibrium single-molecule pulling experiments},\ }\href@noop {}
  {\bibfield  {journal} {\bibinfo  {journal} {Proceedings of the National
  Academy of Sciences}\ }\textbf {\bibinfo {volume} {98}},\ \bibinfo {pages}
  {3658} (\bibinfo {year} {2001})}\BibitemShut {NoStop}%
\bibitem [{\citenamefont {Liphardt}\ \emph {et~al.}(2002)\citenamefont
  {Liphardt}, \citenamefont {Dumont}, \citenamefont {Smith}, \citenamefont
  {Tinoco},\ and\ \citenamefont {Bustamante}}]{liphardt2002equilibrium}%
  \BibitemOpen
  \bibfield  {author} {\bibinfo {author} {\bibfnamefont {J.}~\bibnamefont
  {Liphardt}}, \bibinfo {author} {\bibfnamefont {S.}~\bibnamefont {Dumont}},
  \bibinfo {author} {\bibfnamefont {S.~B.}\ \bibnamefont {Smith}}, \bibinfo
  {author} {\bibfnamefont {I.}~\bibnamefont {Tinoco}},\ and\ \bibinfo {author}
  {\bibfnamefont {C.}~\bibnamefont {Bustamante}},\ }\bibfield  {title}
  {\bibinfo {title} {Equilibrium information from nonequilibrium measurements
  in an experimental test of jarzynski's equality},\ }\href@noop {} {\bibfield
  {journal} {\bibinfo  {journal} {Science}\ }\textbf {\bibinfo {volume}
  {296}},\ \bibinfo {pages} {1832} (\bibinfo {year} {2002})}\BibitemShut
  {NoStop}%
\bibitem [{\citenamefont {Verley}\ \emph
  {et~al.}(2014{\natexlab{a}})\citenamefont {Verley}, \citenamefont {Esposito},
  \citenamefont {Willaert},\ and\ \citenamefont {Van~den
  Broeck}}]{verley2014unlikely}%
  \BibitemOpen
  \bibfield  {author} {\bibinfo {author} {\bibfnamefont {G.}~\bibnamefont
  {Verley}}, \bibinfo {author} {\bibfnamefont {M.}~\bibnamefont {Esposito}},
  \bibinfo {author} {\bibfnamefont {T.}~\bibnamefont {Willaert}},\ and\
  \bibinfo {author} {\bibfnamefont {C.}~\bibnamefont {Van~den Broeck}},\
  }\bibfield  {title} {\bibinfo {title} {The unlikely carnot efficiency},\
  }\href@noop {} {\bibfield  {journal} {\bibinfo  {journal} {Nature
  communications}\ }\textbf {\bibinfo {volume} {5}},\ \bibinfo {pages} {1}
  (\bibinfo {year} {2014}{\natexlab{a}})}\BibitemShut {NoStop}%
\bibitem [{\citenamefont {Verley}\ \emph
  {et~al.}(2014{\natexlab{b}})\citenamefont {Verley}, \citenamefont {Willaert},
  \citenamefont {Van~den Broeck},\ and\ \citenamefont
  {Esposito}}]{verley2014universal}%
  \BibitemOpen
  \bibfield  {author} {\bibinfo {author} {\bibfnamefont {G.}~\bibnamefont
  {Verley}}, \bibinfo {author} {\bibfnamefont {T.}~\bibnamefont {Willaert}},
  \bibinfo {author} {\bibfnamefont {C.}~\bibnamefont {Van~den Broeck}},\ and\
  \bibinfo {author} {\bibfnamefont {M.}~\bibnamefont {Esposito}},\ }\bibfield
  {title} {\bibinfo {title} {Universal theory of efficiency fluctuations},\
  }\href@noop {} {\bibfield  {journal} {\bibinfo  {journal} {Physical Review
  E}\ }\textbf {\bibinfo {volume} {90}},\ \bibinfo {pages} {052145} (\bibinfo
  {year} {2014}{\natexlab{b}})}\BibitemShut {NoStop}%
\bibitem [{\citenamefont {Manikandan}\ \emph {et~al.}(2019)\citenamefont
  {Manikandan}, \citenamefont {Dabelow}, \citenamefont {Eichhorn},\ and\
  \citenamefont {Krishnamurthy}}]{manikandan:EF}%
  \BibitemOpen
  \bibfield  {author} {\bibinfo {author} {\bibfnamefont {S.~K.}\ \bibnamefont
  {Manikandan}}, \bibinfo {author} {\bibfnamefont {L.}~\bibnamefont {Dabelow}},
  \bibinfo {author} {\bibfnamefont {R.}~\bibnamefont {Eichhorn}},\ and\
  \bibinfo {author} {\bibfnamefont {S.}~\bibnamefont {Krishnamurthy}},\
  }\bibfield  {title} {\bibinfo {title} {Efficiency fluctuations in microscopic
  machines},\ }\href {https://doi.org/10.1103/PhysRevLett.122.140601}
  {\bibfield  {journal} {\bibinfo  {journal} {Phys. Rev. Lett.}\ }\textbf
  {\bibinfo {volume} {122}},\ \bibinfo {pages} {140601} (\bibinfo {year}
  {2019})}\BibitemShut {NoStop}%
\bibitem [{\citenamefont {Gingrich}\ \emph {et~al.}(2016)\citenamefont
  {Gingrich}, \citenamefont {Horowitz}, \citenamefont {Perunov},\ and\
  \citenamefont {England}}]{dba}%
  \BibitemOpen
  \bibfield  {author} {\bibinfo {author} {\bibfnamefont {T.~R.}\ \bibnamefont
  {Gingrich}}, \bibinfo {author} {\bibfnamefont {J.~M.}\ \bibnamefont
  {Horowitz}}, \bibinfo {author} {\bibfnamefont {N.}~\bibnamefont {Perunov}},\
  and\ \bibinfo {author} {\bibfnamefont {J.~L.}\ \bibnamefont {England}},\
  }\bibfield  {title} {\bibinfo {title} {Dissipation bounds all steady-state
  current fluctuations},\ }\href
  {https://doi.org/10.1103/PhysRevLett.116.120601} {\bibfield  {journal}
  {\bibinfo  {journal} {Phys. Rev. Lett.}\ }\textbf {\bibinfo {volume} {116}},\
  \bibinfo {pages} {120601} (\bibinfo {year} {2016})}\BibitemShut {NoStop}%
\bibitem [{\citenamefont {Pietzonka}\ \emph {et~al.}(2016)\citenamefont
  {Pietzonka}, \citenamefont {Barato},\ and\ \citenamefont {Seifert}}]{univc}%
  \BibitemOpen
  \bibfield  {author} {\bibinfo {author} {\bibfnamefont {P.}~\bibnamefont
  {Pietzonka}}, \bibinfo {author} {\bibfnamefont {A.~C.}\ \bibnamefont
  {Barato}},\ and\ \bibinfo {author} {\bibfnamefont {U.}~\bibnamefont
  {Seifert}},\ }\bibfield  {title} {\bibinfo {title} {Universal bounds on
  current fluctuations},\ }\href {https://doi.org/10.1103/PhysRevE.93.052145}
  {\bibfield  {journal} {\bibinfo  {journal} {Phys. Rev. E}\ }\textbf {\bibinfo
  {volume} {93}},\ \bibinfo {pages} {052145} (\bibinfo {year}
  {2016})}\BibitemShut {NoStop}%
\bibitem [{\citenamefont {Horowitz}\ and\ \citenamefont
  {Gingrich}(2017)}]{pft}%
  \BibitemOpen
  \bibfield  {author} {\bibinfo {author} {\bibfnamefont {J.~M.}\ \bibnamefont
  {Horowitz}}\ and\ \bibinfo {author} {\bibfnamefont {T.~R.}\ \bibnamefont
  {Gingrich}},\ }\bibfield  {title} {\bibinfo {title} {Proof of the finite-time
  thermodynamic uncertainty relation for steady-state currents},\ }\href
  {https://doi.org/10.1103/PhysRevE.96.020103} {\bibfield  {journal} {\bibinfo
  {journal} {Phys. Rev. E}\ }\textbf {\bibinfo {volume} {96}},\ \bibinfo
  {pages} {020103} (\bibinfo {year} {2017})}\BibitemShut {NoStop}%
\bibitem [{\citenamefont {Barato}\ and\ \citenamefont
  {Seifert}(2015)}]{TURprl}%
  \BibitemOpen
  \bibfield  {author} {\bibinfo {author} {\bibfnamefont {A.~C.}\ \bibnamefont
  {Barato}}\ and\ \bibinfo {author} {\bibfnamefont {U.}~\bibnamefont
  {Seifert}},\ }\bibfield  {title} {\bibinfo {title} {Thermodynamic uncertainty
  relation for biomolecular processes},\ }\href
  {https://doi.org/10.1103/PhysRevLett.114.158101} {\bibfield  {journal}
  {\bibinfo  {journal} {Phys. Rev. Lett.}\ }\textbf {\bibinfo {volume} {114}},\
  \bibinfo {pages} {158101} (\bibinfo {year} {2015})}\BibitemShut {NoStop}%
\bibitem [{\citenamefont {Horowitz}\ and\ \citenamefont
  {Gingrich}(2020)}]{TURnatphys}%
  \BibitemOpen
  \bibfield  {author} {\bibinfo {author} {\bibfnamefont {J.~M.}\ \bibnamefont
  {Horowitz}}\ and\ \bibinfo {author} {\bibfnamefont {T.~R.}\ \bibnamefont
  {Gingrich}},\ }\bibfield  {title} {\bibinfo {title} {Thermodynamic
  uncertainty relations constrain non-equilibrium fluctuations},\ }\href@noop
  {} {\bibfield  {journal} {\bibinfo  {journal} {Nature Physics}\ }\textbf
  {\bibinfo {volume} {16}},\ \bibinfo {pages} {15} (\bibinfo {year}
  {2020})}\BibitemShut {NoStop}%
\bibitem [{\citenamefont {Seifert}(2019)}]{TURannrev}%
  \BibitemOpen
  \bibfield  {author} {\bibinfo {author} {\bibfnamefont {U.}~\bibnamefont
  {Seifert}},\ }\bibfield  {title} {\bibinfo {title} {From stochastic
  thermodynamics to thermodynamic inference},\ }\href@noop {} {\bibfield
  {journal} {\bibinfo  {journal} {Annual Review of Condensed Matter Physics}\
  }\textbf {\bibinfo {volume} {10}},\ \bibinfo {pages} {171} (\bibinfo {year}
  {2019})}\BibitemShut {NoStop}%
\bibitem [{\citenamefont {Ellis}(2006)}]{ellis2006entropy}%
  \BibitemOpen
  \bibfield  {author} {\bibinfo {author} {\bibfnamefont {R.~S.}\ \bibnamefont
  {Ellis}},\ }\href@noop {} {\emph {\bibinfo {title} {Entropy, large
  deviations, and statistical mechanics}}},\ Vol.\ \bibinfo {volume} {1431}\
  (\bibinfo  {publisher} {Taylor \& Francis},\ \bibinfo {year}
  {2006})\BibitemShut {NoStop}%
\bibitem [{\citenamefont {Touchette}(2009)}]{touchette2009large}%
  \BibitemOpen
  \bibfield  {author} {\bibinfo {author} {\bibfnamefont {H.}~\bibnamefont
  {Touchette}},\ }\bibfield  {title} {\bibinfo {title} {The large deviation
  approach to statistical mechanics},\ }\href@noop {} {\bibfield  {journal}
  {\bibinfo  {journal} {Physics Reports}\ }\textbf {\bibinfo {volume} {478}},\
  \bibinfo {pages} {1} (\bibinfo {year} {2009})}\BibitemShut {NoStop}%
\bibitem [{\citenamefont {Lebowitz}\ and\ \citenamefont
  {Spohn}(1999)}]{lebowitz1999gallavotti}%
  \BibitemOpen
  \bibfield  {author} {\bibinfo {author} {\bibfnamefont {J.~L.}\ \bibnamefont
  {Lebowitz}}\ and\ \bibinfo {author} {\bibfnamefont {H.}~\bibnamefont
  {Spohn}},\ }\bibfield  {title} {\bibinfo {title} {A gallavotti--cohen-type
  symmetry in the large deviation functional for stochastic dynamics},\
  }\href@noop {} {\bibfield  {journal} {\bibinfo  {journal} {Journal of
  Statistical Physics}\ }\textbf {\bibinfo {volume} {95}},\ \bibinfo {pages}
  {333} (\bibinfo {year} {1999})}\BibitemShut {NoStop}%
\bibitem [{\citenamefont {Derrida}\ and\ \citenamefont
  {Lebowitz}(1998)}]{BernardD}%
  \BibitemOpen
  \bibfield  {author} {\bibinfo {author} {\bibfnamefont {B.}~\bibnamefont
  {Derrida}}\ and\ \bibinfo {author} {\bibfnamefont {J.~L.}\ \bibnamefont
  {Lebowitz}},\ }\bibfield  {title} {\bibinfo {title} {Exact large deviation
  function in the asymmetric exclusion process},\ }\href
  {https://doi.org/10.1103/PhysRevLett.80.209} {\bibfield  {journal} {\bibinfo
  {journal} {Phys. Rev. Lett.}\ }\textbf {\bibinfo {volume} {80}},\ \bibinfo
  {pages} {209} (\bibinfo {year} {1998})}\BibitemShut {NoStop}%
\bibitem [{\citenamefont {Chetrite}\ and\ \citenamefont
  {Touchette}(2015)}]{chetrite2015nonequilibrium}%
  \BibitemOpen
  \bibfield  {author} {\bibinfo {author} {\bibfnamefont {R.}~\bibnamefont
  {Chetrite}}\ and\ \bibinfo {author} {\bibfnamefont {H.}~\bibnamefont
  {Touchette}},\ }\bibfield  {title} {\bibinfo {title} {Nonequilibrium markov
  processes conditioned on large deviations},\ }in\ \href@noop {} {\emph
  {\bibinfo {booktitle} {Annales Henri Poincar{\'e}}}},\ Vol.~\bibinfo {volume}
  {16}\ (\bibinfo {organization} {Springer},\ \bibinfo {year} {2015})\ pp.\
  \bibinfo {pages} {2005--2057}\BibitemShut {NoStop}%
\bibitem [{\citenamefont {Mehl}\ \emph {et~al.}(2008)\citenamefont {Mehl},
  \citenamefont {Speck},\ and\ \citenamefont {Seifert}}]{mehl2008large}%
  \BibitemOpen
  \bibfield  {author} {\bibinfo {author} {\bibfnamefont {J.}~\bibnamefont
  {Mehl}}, \bibinfo {author} {\bibfnamefont {T.}~\bibnamefont {Speck}},\ and\
  \bibinfo {author} {\bibfnamefont {U.}~\bibnamefont {Seifert}},\ }\bibfield
  {title} {\bibinfo {title} {Large deviation function for entropy production in
  driven one-dimensional systems},\ }\href@noop {} {\bibfield  {journal}
  {\bibinfo  {journal} {Physical Review E}\ }\textbf {\bibinfo {volume} {78}},\
  \bibinfo {pages} {011123} (\bibinfo {year} {2008})}\BibitemShut {NoStop}%
\bibitem [{\citenamefont {Speck}\ \emph {et~al.}(2012)\citenamefont {Speck},
  \citenamefont {Engel},\ and\ \citenamefont {Seifert}}]{speck2012large}%
  \BibitemOpen
  \bibfield  {author} {\bibinfo {author} {\bibfnamefont {T.}~\bibnamefont
  {Speck}}, \bibinfo {author} {\bibfnamefont {A.}~\bibnamefont {Engel}},\ and\
  \bibinfo {author} {\bibfnamefont {U.}~\bibnamefont {Seifert}},\ }\bibfield
  {title} {\bibinfo {title} {The large deviation function for entropy
  production: the optimal trajectory and the role of fluctuations},\
  }\href@noop {} {\bibfield  {journal} {\bibinfo  {journal} {Journal of
  Statistical Mechanics: Theory and Experiment}\ }\textbf {\bibinfo {volume}
  {2012}},\ \bibinfo {pages} {P12001} (\bibinfo {year} {2012})}\BibitemShut
  {NoStop}%
\bibitem [{\citenamefont {Kundu}\ \emph {et~al.}(2011)\citenamefont {Kundu},
  \citenamefont {Sabhapandit},\ and\ \citenamefont {Dhar}}]{kundu2011large}%
  \BibitemOpen
  \bibfield  {author} {\bibinfo {author} {\bibfnamefont {A.}~\bibnamefont
  {Kundu}}, \bibinfo {author} {\bibfnamefont {S.}~\bibnamefont {Sabhapandit}},\
  and\ \bibinfo {author} {\bibfnamefont {A.}~\bibnamefont {Dhar}},\ }\bibfield
  {title} {\bibinfo {title} {Large deviations of heat flow in harmonic
  chains},\ }\href@noop {} {\bibfield  {journal} {\bibinfo  {journal} {Journal
  of Statistical Mechanics: Theory and Experiment}\ }\textbf {\bibinfo {volume}
  {2011}},\ \bibinfo {pages} {P03007} (\bibinfo {year} {2011})}\BibitemShut
  {NoStop}%
\bibitem [{\citenamefont {Sabhapandit}(2012)}]{sabhapandit2012heat}%
  \BibitemOpen
  \bibfield  {author} {\bibinfo {author} {\bibfnamefont {S.}~\bibnamefont
  {Sabhapandit}},\ }\bibfield  {title} {\bibinfo {title} {Heat and work
  fluctuations for a harmonic oscillator},\ }\href@noop {} {\bibfield
  {journal} {\bibinfo  {journal} {Physical Review E}\ }\textbf {\bibinfo
  {volume} {85}},\ \bibinfo {pages} {021108} (\bibinfo {year}
  {2012})}\BibitemShut {NoStop}%
\bibitem [{\citenamefont {Verley}\ \emph
  {et~al.}(2014{\natexlab{c}})\citenamefont {Verley}, \citenamefont {Van~den
  Broeck},\ and\ \citenamefont {Esposito}}]{verley2014work}%
  \BibitemOpen
  \bibfield  {author} {\bibinfo {author} {\bibfnamefont {G.}~\bibnamefont
  {Verley}}, \bibinfo {author} {\bibfnamefont {C.}~\bibnamefont {Van~den
  Broeck}},\ and\ \bibinfo {author} {\bibfnamefont {M.}~\bibnamefont
  {Esposito}},\ }\bibfield  {title} {\bibinfo {title} {Work statistics in
  stochastically driven systems},\ }\href@noop {} {\bibfield  {journal}
  {\bibinfo  {journal} {New Journal of Physics}\ }\textbf {\bibinfo {volume}
  {16}},\ \bibinfo {pages} {095001} (\bibinfo {year}
  {2014}{\natexlab{c}})}\BibitemShut {NoStop}%
\bibitem [{\citenamefont {Morgado}\ and\ \citenamefont
  {Duarte~Queir\'os}(2014)}]{NewRef}%
  \BibitemOpen
  \bibfield  {author} {\bibinfo {author} {\bibfnamefont {W.~A.~M.}\
  \bibnamefont {Morgado}}\ and\ \bibinfo {author} {\bibfnamefont {S.~M.}\
  \bibnamefont {Duarte~Queir\'os}},\ }\bibfield  {title} {\bibinfo {title}
  {Thermostatistics of small nonlinear systems: Gaussian thermal bath},\ }\href
  {https://doi.org/10.1103/PhysRevE.90.022110} {\bibfield  {journal} {\bibinfo
  {journal} {Phys. Rev. E}\ }\textbf {\bibinfo {volume} {90}},\ \bibinfo
  {pages} {022110} (\bibinfo {year} {2014})}\BibitemShut {NoStop}%
\bibitem [{\citenamefont {Saito}\ and\ \citenamefont
  {Dhar}(2016)}]{saito2016waiting}%
  \BibitemOpen
  \bibfield  {author} {\bibinfo {author} {\bibfnamefont {K.}~\bibnamefont
  {Saito}}\ and\ \bibinfo {author} {\bibfnamefont {A.}~\bibnamefont {Dhar}},\
  }\bibfield  {title} {\bibinfo {title} {Waiting for rare entropic
  fluctuations},\ }\href@noop {} {\bibfield  {journal} {\bibinfo  {journal}
  {EPL (Europhysics Letters)}\ }\textbf {\bibinfo {volume} {114}},\ \bibinfo
  {pages} {50004} (\bibinfo {year} {2016})}\BibitemShut {NoStop}%
\bibitem [{\citenamefont {Garrahan}(2017)}]{simplebounds}%
  \BibitemOpen
  \bibfield  {author} {\bibinfo {author} {\bibfnamefont {J.~P.}\ \bibnamefont
  {Garrahan}},\ }\bibfield  {title} {\bibinfo {title} {Simple bounds on
  fluctuations and uncertainty relations for first-passage times of counting
  observables},\ }\href {https://doi.org/10.1103/PhysRevE.95.032134} {\bibfield
   {journal} {\bibinfo  {journal} {Phys. Rev. E}\ }\textbf {\bibinfo {volume}
  {95}},\ \bibinfo {pages} {032134} (\bibinfo {year} {2017})}\BibitemShut
  {NoStop}%
\bibitem [{\citenamefont {Singh}\ and\ \citenamefont
  {Kundu}(2019)}]{singh2019generalised}%
  \BibitemOpen
  \bibfield  {author} {\bibinfo {author} {\bibfnamefont {P.}~\bibnamefont
  {Singh}}\ and\ \bibinfo {author} {\bibfnamefont {A.}~\bibnamefont {Kundu}},\
  }\bibfield  {title} {\bibinfo {title} {Generalised ‘arcsine’laws for
  run-and-tumble particle in one dimension},\ }\href@noop {} {\bibfield
  {journal} {\bibinfo  {journal} {Journal of Statistical Mechanics: Theory and
  Experiment}\ }\textbf {\bibinfo {volume} {2019}},\ \bibinfo {pages} {083205}
  (\bibinfo {year} {2019})}\BibitemShut {NoStop}%
\bibitem [{\citenamefont {Bodineau}\ and\ \citenamefont
  {Derrida}(2004)}]{additivity}%
  \BibitemOpen
  \bibfield  {author} {\bibinfo {author} {\bibfnamefont {T.}~\bibnamefont
  {Bodineau}}\ and\ \bibinfo {author} {\bibfnamefont {B.}~\bibnamefont
  {Derrida}},\ }\bibfield  {title} {\bibinfo {title} {Current fluctuations in
  nonequilibrium diffusive systems: An additivity principle},\ }\href
  {https://doi.org/10.1103/PhysRevLett.92.180601} {\bibfield  {journal}
  {\bibinfo  {journal} {Phys. Rev. Lett.}\ }\textbf {\bibinfo {volume} {92}},\
  \bibinfo {pages} {180601} (\bibinfo {year} {2004})}\BibitemShut {NoStop}%
\bibitem [{\citenamefont {Derrida}(2007)}]{derrida2007non}%
  \BibitemOpen
  \bibfield  {author} {\bibinfo {author} {\bibfnamefont {B.}~\bibnamefont
  {Derrida}},\ }\bibfield  {title} {\bibinfo {title} {Non-equilibrium steady
  states: fluctuations and large deviations of the density and of the
  current},\ }\href@noop {} {\bibfield  {journal} {\bibinfo  {journal} {Journal
  of Statistical Mechanics: Theory and Experiment}\ }\textbf {\bibinfo {volume}
  {2007}},\ \bibinfo {pages} {P07023} (\bibinfo {year} {2007})}\BibitemShut
  {NoStop}%
\bibitem [{\citenamefont {Barato}\ \emph {et~al.}(2018)\citenamefont {Barato},
  \citenamefont {Rold{\'a}n}, \citenamefont {Mart{\'\i}nez},\ and\
  \citenamefont {Pigolotti}}]{barato2018arcsine}%
  \BibitemOpen
  \bibfield  {author} {\bibinfo {author} {\bibfnamefont {A.~C.}\ \bibnamefont
  {Barato}}, \bibinfo {author} {\bibfnamefont {{\'E}.}~\bibnamefont
  {Rold{\'a}n}}, \bibinfo {author} {\bibfnamefont {I.~A.}\ \bibnamefont
  {Mart{\'\i}nez}},\ and\ \bibinfo {author} {\bibfnamefont {S.}~\bibnamefont
  {Pigolotti}},\ }\bibfield  {title} {\bibinfo {title} {Arcsine laws in
  stochastic thermodynamics},\ }\href@noop {} {\bibfield  {journal} {\bibinfo
  {journal} {Physical review letters}\ }\textbf {\bibinfo {volume} {121}},\
  \bibinfo {pages} {090601} (\bibinfo {year} {2018})}\BibitemShut {NoStop}%
\bibitem [{\citenamefont {L{\'e}vy}(1940)}]{levy1940certains}%
  \BibitemOpen
  \bibfield  {author} {\bibinfo {author} {\bibfnamefont {P.}~\bibnamefont
  {L{\'e}vy}},\ }\bibfield  {title} {\bibinfo {title} {On some homogeneous
  stochastic processes},\ }\href@noop {} {\bibfield  {journal} {\bibinfo
  {journal} {Compositio mathematica}\ }\textbf {\bibinfo {volume} {7}},\
  \bibinfo {pages} {283} (\bibinfo {year} {1940})}\BibitemShut {NoStop}%
\bibitem [{\citenamefont {Manikandan}\ \emph {et~al.}(2020)\citenamefont
  {Manikandan}, \citenamefont {Gupta},\ and\ \citenamefont
  {Krishnamurthy}}]{manikandan2020inferring}%
  \BibitemOpen
  \bibfield  {author} {\bibinfo {author} {\bibfnamefont {S.~K.}\ \bibnamefont
  {Manikandan}}, \bibinfo {author} {\bibfnamefont {D.}~\bibnamefont {Gupta}},\
  and\ \bibinfo {author} {\bibfnamefont {S.}~\bibnamefont {Krishnamurthy}},\
  }\bibfield  {title} {\bibinfo {title} {Inferring entropy production from
  short experiments},\ }\href@noop {} {\bibfield  {journal} {\bibinfo
  {journal} {Physical review letters}\ }\textbf {\bibinfo {volume} {124}},\
  \bibinfo {pages} {120603} (\bibinfo {year} {2020})}\BibitemShut {NoStop}%
\bibitem [{\citenamefont {Otsubo}\ \emph
  {et~al.}(2020{\natexlab{a}})\citenamefont {Otsubo}, \citenamefont {Ito},
  \citenamefont {Dechant},\ and\ \citenamefont {Sagawa}}]{shun}%
  \BibitemOpen
  \bibfield  {author} {\bibinfo {author} {\bibfnamefont {S.}~\bibnamefont
  {Otsubo}}, \bibinfo {author} {\bibfnamefont {S.}~\bibnamefont {Ito}},
  \bibinfo {author} {\bibfnamefont {A.}~\bibnamefont {Dechant}},\ and\ \bibinfo
  {author} {\bibfnamefont {T.}~\bibnamefont {Sagawa}},\ }\bibfield  {title}
  {\bibinfo {title} {Estimating entropy production by machine learning of
  short-time fluctuating currents},\ }\href
  {https://doi.org/10.1103/PhysRevE.101.062106} {\bibfield  {journal} {\bibinfo
   {journal} {Phys. Rev. E}\ }\textbf {\bibinfo {volume} {101}},\ \bibinfo
  {pages} {062106} (\bibinfo {year} {2020}{\natexlab{a}})}\BibitemShut
  {NoStop}%
\bibitem [{\citenamefont {Van~Vu}\ \emph {et~al.}(2020)\citenamefont {Van~Vu},
  \citenamefont {Vo},\ and\ \citenamefont {Hasegawa}}]{tan}%
  \BibitemOpen
  \bibfield  {author} {\bibinfo {author} {\bibfnamefont {T.}~\bibnamefont
  {Van~Vu}}, \bibinfo {author} {\bibfnamefont {V.~T.}\ \bibnamefont {Vo}},\
  and\ \bibinfo {author} {\bibfnamefont {Y.}~\bibnamefont {Hasegawa}},\
  }\bibfield  {title} {\bibinfo {title} {Entropy production estimation with
  optimal current},\ }\href {https://doi.org/10.1103/PhysRevE.101.042138}
  {\bibfield  {journal} {\bibinfo  {journal} {Phys. Rev. E}\ }\textbf {\bibinfo
  {volume} {101}},\ \bibinfo {pages} {042138} (\bibinfo {year}
  {2020})}\BibitemShut {NoStop}%
\bibitem [{\citenamefont {Manikandan}\ \emph {et~al.}(2021)\citenamefont
  {Manikandan}, \citenamefont {Ghosh}, \citenamefont {Kundu}, \citenamefont
  {Das}, \citenamefont {Agrawal}, \citenamefont {Mitra}, \citenamefont
  {Banerjee},\ and\ \citenamefont
  {Krishnamurthy}}]{manikandan2021quantitative}%
  \BibitemOpen
  \bibfield  {author} {\bibinfo {author} {\bibfnamefont {S.~K.}\ \bibnamefont
  {Manikandan}}, \bibinfo {author} {\bibfnamefont {S.}~\bibnamefont {Ghosh}},
  \bibinfo {author} {\bibfnamefont {A.}~\bibnamefont {Kundu}}, \bibinfo
  {author} {\bibfnamefont {B.}~\bibnamefont {Das}}, \bibinfo {author}
  {\bibfnamefont {V.}~\bibnamefont {Agrawal}}, \bibinfo {author} {\bibfnamefont
  {D.}~\bibnamefont {Mitra}}, \bibinfo {author} {\bibfnamefont
  {A.}~\bibnamefont {Banerjee}},\ and\ \bibinfo {author} {\bibfnamefont
  {S.}~\bibnamefont {Krishnamurthy}},\ }\bibfield  {title} {\bibinfo {title}
  {Quantitative analysis of non-equilibrium systems from short-time
  experimental data},\ }\href
  {https://www.nature.com/articles/s42005-021-00766-2} {\bibfield  {journal}
  {\bibinfo  {journal} {Communications Physics}\ }\textbf {\bibinfo {volume}
  {4}},\ \bibinfo {pages} {1} (\bibinfo {year} {2021})}\BibitemShut {NoStop}%
\bibitem [{\citenamefont {Otsubo}\ \emph
  {et~al.}(2020{\natexlab{b}})\citenamefont {Otsubo}, \citenamefont
  {Manikandan}, \citenamefont {Sagawa},\ and\ \citenamefont
  {Krishnamurthy}}]{otsubo2020estimating}%
  \BibitemOpen
  \bibfield  {author} {\bibinfo {author} {\bibfnamefont {S.}~\bibnamefont
  {Otsubo}}, \bibinfo {author} {\bibfnamefont {S.~K.}\ \bibnamefont
  {Manikandan}}, \bibinfo {author} {\bibfnamefont {T.}~\bibnamefont {Sagawa}},\
  and\ \bibinfo {author} {\bibfnamefont {S.}~\bibnamefont {Krishnamurthy}},\
  }\bibfield  {title} {\bibinfo {title} {Estimating entropy production along a
  single non-equilibrium trajectory},\ }\href@noop {} {\bibfield  {journal}
  {\bibinfo  {journal} {arXiv preprint arXiv:2010.03852}\ } (\bibinfo {year}
  {2020}{\natexlab{b}})}\BibitemShut {NoStop}%
\bibitem [{\citenamefont {Jack}(2020)}]{jack2020ergodicity}%
  \BibitemOpen
  \bibfield  {author} {\bibinfo {author} {\bibfnamefont {R.~L.}\ \bibnamefont
  {Jack}},\ }\bibfield  {title} {\bibinfo {title} {Ergodicity and large
  deviations in physical systems with stochastic dynamics},\ }\href@noop {}
  {\bibfield  {journal} {\bibinfo  {journal} {The European Physical Journal B}\
  }\textbf {\bibinfo {volume} {93}},\ \bibinfo {pages} {1} (\bibinfo {year}
  {2020})}\BibitemShut {NoStop}%
\bibitem [{\citenamefont {Seifert}(2005)}]{seifertep}%
  \BibitemOpen
  \bibfield  {author} {\bibinfo {author} {\bibfnamefont {U.}~\bibnamefont
  {Seifert}},\ }\bibfield  {title} {\bibinfo {title} {Entropy production along
  a stochastic trajectory and an integral fluctuation theorem},\ }\href
  {https://doi.org/10.1103/PhysRevLett.95.040602} {\bibfield  {journal}
  {\bibinfo  {journal} {Phys. Rev. Lett.}\ }\textbf {\bibinfo {volume} {95}},\
  \bibinfo {pages} {040602} (\bibinfo {year} {2005})}\BibitemShut {NoStop}%
\bibitem [{\citenamefont {Merhav}\ and\ \citenamefont
  {Kafri}(2010)}]{merhav2010statistical}%
  \BibitemOpen
  \bibfield  {author} {\bibinfo {author} {\bibfnamefont {N.}~\bibnamefont
  {Merhav}}\ and\ \bibinfo {author} {\bibfnamefont {Y.}~\bibnamefont {Kafri}},\
  }\bibfield  {title} {\bibinfo {title} {Statistical properties of entropy
  production derived from fluctuation theorems},\ }\href@noop {} {\bibfield
  {journal} {\bibinfo  {journal} {Journal of Statistical Mechanics: Theory and
  Experiment}\ }\textbf {\bibinfo {volume} {2010}},\ \bibinfo {pages} {P12022}
  (\bibinfo {year} {2010})}\BibitemShut {NoStop}%
\bibitem [{\citenamefont {Pietzonka}\ \emph {et~al.}(2017)\citenamefont
  {Pietzonka}, \citenamefont {Ritort},\ and\ \citenamefont {Seifert}}]{ftg}%
  \BibitemOpen
  \bibfield  {author} {\bibinfo {author} {\bibfnamefont {P.}~\bibnamefont
  {Pietzonka}}, \bibinfo {author} {\bibfnamefont {F.}~\bibnamefont {Ritort}},\
  and\ \bibinfo {author} {\bibfnamefont {U.}~\bibnamefont {Seifert}},\
  }\bibfield  {title} {\bibinfo {title} {Finite-time generalization of the
  thermodynamic uncertainty relation},\ }\href
  {https://doi.org/10.1103/PhysRevE.96.012101} {\bibfield  {journal} {\bibinfo
  {journal} {Phys. Rev. E}\ }\textbf {\bibinfo {volume} {96}},\ \bibinfo
  {pages} {012101} (\bibinfo {year} {2017})}\BibitemShut {NoStop}%
\bibitem [{\citenamefont {Dechant}\ and\ \citenamefont
  {Sasa}(2018)}]{dechant2018current}%
  \BibitemOpen
  \bibfield  {author} {\bibinfo {author} {\bibfnamefont {A.}~\bibnamefont
  {Dechant}}\ and\ \bibinfo {author} {\bibfnamefont {S.-i.}\ \bibnamefont
  {Sasa}},\ }\bibfield  {title} {\bibinfo {title} {Current fluctuations and
  transport efficiency for general langevin systems},\ }\href@noop {}
  {\bibfield  {journal} {\bibinfo  {journal} {Journal of Statistical Mechanics:
  Theory and Experiment}\ }\textbf {\bibinfo {volume} {2018}},\ \bibinfo
  {pages} {063209} (\bibinfo {year} {2018})}\BibitemShut {NoStop}%
\bibitem [{\citenamefont {Hasegawa}\ and\ \citenamefont
  {Van~Vu}(2019)}]{flcunc}%
  \BibitemOpen
  \bibfield  {author} {\bibinfo {author} {\bibfnamefont {Y.}~\bibnamefont
  {Hasegawa}}\ and\ \bibinfo {author} {\bibfnamefont {T.}~\bibnamefont
  {Van~Vu}},\ }\bibfield  {title} {\bibinfo {title} {Fluctuation theorem
  uncertainty relation},\ }\href
  {https://doi.org/10.1103/PhysRevLett.123.110602} {\bibfield  {journal}
  {\bibinfo  {journal} {Phys. Rev. Lett.}\ }\textbf {\bibinfo {volume} {123}},\
  \bibinfo {pages} {110602} (\bibinfo {year} {2019})}\BibitemShut {NoStop}%
\bibitem [{\citenamefont {Neri}\ \emph {et~al.}(2017)\citenamefont {Neri},
  \citenamefont {Rold\'an},\ and\ \citenamefont {J\"ulicher}}]{neri}%
  \BibitemOpen
  \bibfield  {author} {\bibinfo {author} {\bibfnamefont {I.}~\bibnamefont
  {Neri}}, \bibinfo {author} {\bibfnamefont {E.}~\bibnamefont {Rold\'an}},\
  and\ \bibinfo {author} {\bibfnamefont {F.}~\bibnamefont {J\"ulicher}},\
  }\bibfield  {title} {\bibinfo {title} {Statistics of infima and stopping
  times of entropy production and applications to active molecular processes},\
  }\href {https://doi.org/10.1103/PhysRevX.7.011019} {\bibfield  {journal}
  {\bibinfo  {journal} {Phys. Rev. X}\ }\textbf {\bibinfo {volume} {7}},\
  \bibinfo {pages} {011019} (\bibinfo {year} {2017})}\BibitemShut {NoStop}%
\bibitem [{\citenamefont {Pigolotti}\ \emph {et~al.}(2017)\citenamefont
  {Pigolotti}, \citenamefont {Neri}, \citenamefont {Rold\'an},\ and\
  \citenamefont {J\"ulicher}}]{generic}%
  \BibitemOpen
  \bibfield  {author} {\bibinfo {author} {\bibfnamefont {S.}~\bibnamefont
  {Pigolotti}}, \bibinfo {author} {\bibfnamefont {I.}~\bibnamefont {Neri}},
  \bibinfo {author} {\bibfnamefont {E.}~\bibnamefont {Rold\'an}},\ and\
  \bibinfo {author} {\bibfnamefont {F.}~\bibnamefont {J\"ulicher}},\ }\bibfield
   {title} {\bibinfo {title} {Generic properties of stochastic entropy
  production},\ }\href {https://doi.org/10.1103/PhysRevLett.119.140604}
  {\bibfield  {journal} {\bibinfo  {journal} {Phys. Rev. Lett.}\ }\textbf
  {\bibinfo {volume} {119}},\ \bibinfo {pages} {140604} (\bibinfo {year}
  {2017})}\BibitemShut {NoStop}%
\bibitem [{\citenamefont {Mori}\ \emph {et~al.}(2021)\citenamefont {Mori},
  \citenamefont {Majumdar},\ and\ \citenamefont
  {Schehr}}]{mori2021distribution}%
  \BibitemOpen
  \bibfield  {author} {\bibinfo {author} {\bibfnamefont {F.}~\bibnamefont
  {Mori}}, \bibinfo {author} {\bibfnamefont {S.~N.}\ \bibnamefont {Majumdar}},\
  and\ \bibinfo {author} {\bibfnamefont {G.}~\bibnamefont {Schehr}},\
  }\bibfield  {title} {\bibinfo {title} {Distribution of the time of the
  maximum for stationary processes},\ }\href@noop {} {\bibfield  {journal}
  {\bibinfo  {journal} {EPL (Europhysics Letters)}\ }\textbf {\bibinfo {volume}
  {135}},\ \bibinfo {pages} {30003} (\bibinfo {year} {2021})}\BibitemShut
  {NoStop}%
\bibitem [{\citenamefont {Gingrich}\ and\ \citenamefont
  {Horowitz}(2017)}]{firstpassage}%
  \BibitemOpen
  \bibfield  {author} {\bibinfo {author} {\bibfnamefont {T.~R.}\ \bibnamefont
  {Gingrich}}\ and\ \bibinfo {author} {\bibfnamefont {J.~M.}\ \bibnamefont
  {Horowitz}},\ }\bibfield  {title} {\bibinfo {title} {Fundamental bounds on
  first passage time fluctuations for currents},\ }\href
  {https://doi.org/10.1103/PhysRevLett.119.170601} {\bibfield  {journal}
  {\bibinfo  {journal} {Phys. Rev. Lett.}\ }\textbf {\bibinfo {volume} {119}},\
  \bibinfo {pages} {170601} (\bibinfo {year} {2017})}\BibitemShut {NoStop}%
\bibitem [{Note1()}]{Note1}%
  \BibitemOpen
  \bibinfo {note} {Note Marcinkiewicz theorem \cite
  {marcinkiewicz1939propriete}, which implies that if a distribution is not
  Gaussian, then it will have further non-zero cumulants.}\BibitemShut {Stop}%
\bibitem [{\citenamefont {Gomez-Solano}\ \emph {et~al.}(2010)\citenamefont
  {Gomez-Solano}, \citenamefont {Bellon}, \citenamefont {Petrosyan},\ and\
  \citenamefont {Ciliberto}}]{gomez:ssw}%
  \BibitemOpen
  \bibfield  {author} {\bibinfo {author} {\bibfnamefont {J.~R.}\ \bibnamefont
  {Gomez-Solano}}, \bibinfo {author} {\bibfnamefont {L.}~\bibnamefont
  {Bellon}}, \bibinfo {author} {\bibfnamefont {A.}~\bibnamefont {Petrosyan}},\
  and\ \bibinfo {author} {\bibfnamefont {S.}~\bibnamefont {Ciliberto}},\
  }\bibfield  {title} {\bibinfo {title} {Steady-state fluctuation relations for
  systems driven by an external random force},\ }\href@noop {} {\bibfield
  {journal} {\bibinfo  {journal} {EPL (Europhysics Letters)}\ }\textbf
  {\bibinfo {volume} {89}},\ \bibinfo {pages} {60003} (\bibinfo {year}
  {2010})}\BibitemShut {NoStop}%
\bibitem [{\citenamefont {Pal}\ and\ \citenamefont
  {Sabhapandit}(2013)}]{Pal:2013wfb}%
  \BibitemOpen
  \bibfield  {author} {\bibinfo {author} {\bibfnamefont {A.}~\bibnamefont
  {Pal}}\ and\ \bibinfo {author} {\bibfnamefont {S.}~\bibnamefont
  {Sabhapandit}},\ }\bibfield  {title} {\bibinfo {title} {Work fluctuations for
  a brownian particle in a harmonic trap with fluctuating locations},\ }\href
  {https://doi.org/10.1103/PhysRevE.87.022138} {\bibfield  {journal} {\bibinfo
  {journal} {Phys. Rev. E}\ }\textbf {\bibinfo {volume} {87}},\ \bibinfo
  {pages} {022138} (\bibinfo {year} {2013})}\BibitemShut {NoStop}%
\bibitem [{\citenamefont {Manikandan}\ and\ \citenamefont
  {Krishnamurthy}(2017)}]{Manikandan:2017awd}%
  \BibitemOpen
  \bibfield  {author} {\bibinfo {author} {\bibfnamefont {S.~K.}\ \bibnamefont
  {Manikandan}}\ and\ \bibinfo {author} {\bibfnamefont {S.}~\bibnamefont
  {Krishnamurthy}},\ }\bibfield  {title} {\bibinfo {title} {Asymptotics of work
  distributions in a stochastically driven system},\ }\href
  {https://doi.org/10.1140/epjb/e2017-80432-9} {\bibfield  {journal} {\bibinfo
  {journal} {The European Physical Journal B}\ }\textbf {\bibinfo {volume}
  {90}},\ \bibinfo {pages} {258} (\bibinfo {year} {2017})}\BibitemShut
  {NoStop}%
\bibitem [{\citenamefont {Manikandan}\ and\ \citenamefont
  {Krishnamurthy}(2018)}]{manikandan2018exact}%
  \BibitemOpen
  \bibfield  {author} {\bibinfo {author} {\bibfnamefont {S.~K.}\ \bibnamefont
  {Manikandan}}\ and\ \bibinfo {author} {\bibfnamefont {S.}~\bibnamefont
  {Krishnamurthy}},\ }\bibfield  {title} {\bibinfo {title} {Exact results for
  the finite time thermodynamic uncertainty relation},\ }\href@noop {}
  {\bibfield  {journal} {\bibinfo  {journal} {Journal of Physics A:
  Mathematical and Theoretical}\ }\textbf {\bibinfo {volume} {51}},\ \bibinfo
  {pages} {11LT01} (\bibinfo {year} {2018})}\BibitemShut {NoStop}%
\bibitem [{Note2()}]{Note2}%
  \BibitemOpen
  \bibinfo {note} {See Eq.\ (14b) in Ref.\ \cite {manikandan2021quantitative}
  for the choice of $c_i$'s that define $J = \Delta S_{tot}$}\BibitemShut
  {NoStop}%
\bibitem [{Note3()}]{Note3}%
  \BibitemOpen
  \bibinfo {note} {The exact analytic calculation involves the computation of
  the finite-time moment generating function $G(\lambda ) = \langle e^{-\lambda
  \Delta S_{tot}}\rangle _t$ using a path integral technique which was
  developed in \cite {Manikandan:2017awd} and used in \cite
  {manikandan2018exact}. For completeness, we reproduce the results (from \cite
  {manikandan2018exact})in the supplemental material.}\BibitemShut {Stop}%
\bibitem [{\citenamefont {Bera}\ \emph {et~al.}(2017)\citenamefont {Bera},
  \citenamefont {Paul}, \citenamefont {Singh}, \citenamefont {Ghosh},
  \citenamefont {Kundu},\ and\ \citenamefont {Banerjee}}]{bera2017scirep}%
  \BibitemOpen
  \bibfield  {author} {\bibinfo {author} {\bibfnamefont {S.}~\bibnamefont
  {Bera}}, \bibinfo {author} {\bibfnamefont {S.}~\bibnamefont {Paul}}, \bibinfo
  {author} {\bibfnamefont {R.}~\bibnamefont {Singh}}, \bibinfo {author}
  {\bibfnamefont {D.}~\bibnamefont {Ghosh}}, \bibinfo {author} {\bibfnamefont
  {A.}~\bibnamefont {Kundu}},\ and\ \bibinfo {author} {\bibfnamefont
  {R.}~\bibnamefont {Banerjee}, \bibfnamefont {A.~\&~Adhikari}},\ }\bibfield
  {title} {\bibinfo {title} {Fast bayesian inference of optical trap stiffness
  and particle diffusion},\ }\href@noop {} {\bibfield  {journal} {\bibinfo
  {journal} {Scientific Reports}\ }\textbf {\bibinfo {volume} {7}},\ \bibinfo
  {pages} {41638} (\bibinfo {year} {2017})}\BibitemShut {NoStop}%
\bibitem [{Note4()}]{Note4}%
  \BibitemOpen
  \bibinfo {note} {Parameters used in the plots are given in section IV of
  supplemental material \cite {supp}}\BibitemShut {NoStop}%
\bibitem [{\citenamefont {Das}\ \emph {et~al.}(2022)\citenamefont {Das},
  \citenamefont {Manikandan},\ and\ \citenamefont
  {Banerjee}}]{das2022inferring}%
  \BibitemOpen
  \bibfield  {author} {\bibinfo {author} {\bibfnamefont {B.}~\bibnamefont
  {Das}}, \bibinfo {author} {\bibfnamefont {S.~K.}\ \bibnamefont
  {Manikandan}},\ and\ \bibinfo {author} {\bibfnamefont {A.}~\bibnamefont
  {Banerjee}},\ }\bibfield  {title} {\bibinfo {title} {Inferring entropy
  production in anharmonic brownian gyrators},\ }\href
  {https://arxiv.org/abs/2204.09283} {\bibfield  {journal} {\bibinfo  {journal}
  {arXiv preprint arXiv:2204.09283}\ } (\bibinfo {year} {2022})}\BibitemShut
  {NoStop}%
\bibitem [{sup()}]{supp}%
  \BibitemOpen
  \href@noop {} {}\bibinfo {howpublished}
  {\url{URL_will_be_inserted_by_publisher}}\BibitemShut {NoStop}%
\bibitem [{\citenamefont {Scovil}\ and\ \citenamefont
  {Schulz-DuBois}(1959)}]{PhysRevLett.2.262}%
  \BibitemOpen
  \bibfield  {author} {\bibinfo {author} {\bibfnamefont {H.~E.~D.}\
  \bibnamefont {Scovil}}\ and\ \bibinfo {author} {\bibfnamefont {E.~O.}\
  \bibnamefont {Schulz-DuBois}},\ }\bibfield  {title} {\bibinfo {title}
  {Three-level masers as heat engines},\ }\href
  {https://doi.org/10.1103/PhysRevLett.2.262} {\bibfield  {journal} {\bibinfo
  {journal} {Phys. Rev. Lett.}\ }\textbf {\bibinfo {volume} {2}},\ \bibinfo
  {pages} {262} (\bibinfo {year} {1959})}\BibitemShut {NoStop}%
\bibitem [{\citenamefont {Zou}\ \emph {et~al.}(2017)\citenamefont {Zou},
  \citenamefont {Jiang}, \citenamefont {Mei}, \citenamefont {Guo},\ and\
  \citenamefont {Du}}]{PhysRevLett.119.050602}%
  \BibitemOpen
  \bibfield  {author} {\bibinfo {author} {\bibfnamefont {Y.}~\bibnamefont
  {Zou}}, \bibinfo {author} {\bibfnamefont {Y.}~\bibnamefont {Jiang}}, \bibinfo
  {author} {\bibfnamefont {Y.}~\bibnamefont {Mei}}, \bibinfo {author}
  {\bibfnamefont {X.}~\bibnamefont {Guo}},\ and\ \bibinfo {author}
  {\bibfnamefont {S.}~\bibnamefont {Du}},\ }\bibfield  {title} {\bibinfo
  {title} {Quantum heat engine using electromagnetically induced
  transparency},\ }\href {https://doi.org/10.1103/PhysRevLett.119.050602}
  {\bibfield  {journal} {\bibinfo  {journal} {Phys. Rev. Lett.}\ }\textbf
  {\bibinfo {volume} {119}},\ \bibinfo {pages} {050602} (\bibinfo {year}
  {2017})}\BibitemShut {NoStop}%
\bibitem [{\citenamefont {Klatzow}\ \emph {et~al.}(2019)\citenamefont
  {Klatzow}, \citenamefont {Becker}, \citenamefont {Ledingham}, \citenamefont
  {Weinzetl}, \citenamefont {Kaczmarek}, \citenamefont {Saunders},
  \citenamefont {Nunn}, \citenamefont {Walmsley}, \citenamefont {Uzdin},\ and\
  \citenamefont {Poem}}]{PhysRevLett.122.110601}%
  \BibitemOpen
  \bibfield  {author} {\bibinfo {author} {\bibfnamefont {J.}~\bibnamefont
  {Klatzow}}, \bibinfo {author} {\bibfnamefont {J.~N.}\ \bibnamefont {Becker}},
  \bibinfo {author} {\bibfnamefont {P.~M.}\ \bibnamefont {Ledingham}}, \bibinfo
  {author} {\bibfnamefont {C.}~\bibnamefont {Weinzetl}}, \bibinfo {author}
  {\bibfnamefont {K.~T.}\ \bibnamefont {Kaczmarek}}, \bibinfo {author}
  {\bibfnamefont {D.~J.}\ \bibnamefont {Saunders}}, \bibinfo {author}
  {\bibfnamefont {J.}~\bibnamefont {Nunn}}, \bibinfo {author} {\bibfnamefont
  {I.~A.}\ \bibnamefont {Walmsley}}, \bibinfo {author} {\bibfnamefont
  {R.}~\bibnamefont {Uzdin}},\ and\ \bibinfo {author} {\bibfnamefont
  {E.}~\bibnamefont {Poem}},\ }\bibfield  {title} {\bibinfo {title}
  {Experimental demonstration of quantum effects in the operation of
  microscopic heat engines},\ }\href
  {https://doi.org/10.1103/PhysRevLett.122.110601} {\bibfield  {journal}
  {\bibinfo  {journal} {Phys. Rev. Lett.}\ }\textbf {\bibinfo {volume} {122}},\
  \bibinfo {pages} {110601} (\bibinfo {year} {2019})}\BibitemShut {NoStop}%
\bibitem [{\citenamefont {Van~Vu}\ and\ \citenamefont
  {Saito}(2022)}]{PhysRevLett.128.140602}%
  \BibitemOpen
  \bibfield  {author} {\bibinfo {author} {\bibfnamefont {T.}~\bibnamefont
  {Van~Vu}}\ and\ \bibinfo {author} {\bibfnamefont {K.}~\bibnamefont {Saito}},\
  }\bibfield  {title} {\bibinfo {title} {Thermodynamics of precision in
  markovian open quantum dynamics},\ }\href
  {https://doi.org/10.1103/PhysRevLett.128.140602} {\bibfield  {journal}
  {\bibinfo  {journal} {Phys. Rev. Lett.}\ }\textbf {\bibinfo {volume} {128}},\
  \bibinfo {pages} {140602} (\bibinfo {year} {2022})}\BibitemShut {NoStop}%
\bibitem [{\citenamefont {Singh}(2020)}]{singh2020optimal}%
  \BibitemOpen
  \bibfield  {author} {\bibinfo {author} {\bibfnamefont {V.}~\bibnamefont
  {Singh}},\ }\bibfield  {title} {\bibinfo {title} {Optimal operation of a
  three-level quantum heat engine and universal nature of efficiency},\
  }\href@noop {} {\bibfield  {journal} {\bibinfo  {journal} {Physical Review
  Research}\ }\textbf {\bibinfo {volume} {2}},\ \bibinfo {pages} {043187}
  (\bibinfo {year} {2020})}\BibitemShut {NoStop}%
\bibitem [{\citenamefont {Manikandan}(2021)}]{PhysRevResearch.3.043108}%
  \BibitemOpen
  \bibfield  {author} {\bibinfo {author} {\bibfnamefont {S.~K.}\ \bibnamefont
  {Manikandan}},\ }\bibfield  {title} {\bibinfo {title} {Equidistant quenches
  in few-level quantum systems},\ }\href
  {https://doi.org/10.1103/PhysRevResearch.3.043108} {\bibfield  {journal}
  {\bibinfo  {journal} {Phys. Rev. Research}\ }\textbf {\bibinfo {volume}
  {3}},\ \bibinfo {pages} {043108} (\bibinfo {year} {2021})}\BibitemShut
  {NoStop}%
\bibitem [{\citenamefont {Milburn}(2020)}]{milburn2020thermodynamics}%
  \BibitemOpen
  \bibfield  {author} {\bibinfo {author} {\bibfnamefont {G.}~\bibnamefont
  {Milburn}},\ }\bibfield  {title} {\bibinfo {title} {The thermodynamics of
  clocks},\ }\href@noop {} {\bibfield  {journal} {\bibinfo  {journal}
  {Contemporary Physics}\ }\textbf {\bibinfo {volume} {61}},\ \bibinfo {pages}
  {69} (\bibinfo {year} {2020})}\BibitemShut {NoStop}%
\bibitem [{Note5()}]{Note5}%
  \BibitemOpen
  \bibinfo {note} {See discussion around Eq. (92) in \cite
  {milburn2020thermodynamics}}\BibitemShut {NoStop}%
\bibitem [{\citenamefont {Vale}\ and\ \citenamefont
  {Fletterick}(1997)}]{vale1997design}%
  \BibitemOpen
  \bibfield  {author} {\bibinfo {author} {\bibfnamefont {R.~D.}\ \bibnamefont
  {Vale}}\ and\ \bibinfo {author} {\bibfnamefont {R.~J.}\ \bibnamefont
  {Fletterick}},\ }\bibfield  {title} {\bibinfo {title} {The design plan of
  kinesin motors},\ }\href@noop {} {\bibfield  {journal} {\bibinfo  {journal}
  {Annual review of cell and developmental biology}\ }\textbf {\bibinfo
  {volume} {13}},\ \bibinfo {pages} {745} (\bibinfo {year} {1997})}\BibitemShut
  {NoStop}%
\bibitem [{\citenamefont {Schnitzer}\ and\ \citenamefont
  {Block}(1997)}]{schnitzer1997kinesin}%
  \BibitemOpen
  \bibfield  {author} {\bibinfo {author} {\bibfnamefont {M.~J.}\ \bibnamefont
  {Schnitzer}}\ and\ \bibinfo {author} {\bibfnamefont {S.~M.}\ \bibnamefont
  {Block}},\ }\bibfield  {title} {\bibinfo {title} {Kinesin hydrolyses one atp
  per 8-nm step},\ }\href@noop {} {\bibfield  {journal} {\bibinfo  {journal}
  {Nature}\ }\textbf {\bibinfo {volume} {388}},\ \bibinfo {pages} {386}
  (\bibinfo {year} {1997})}\BibitemShut {NoStop}%
\bibitem [{\citenamefont {Ariga}\ \emph {et~al.}(2018)\citenamefont {Ariga},
  \citenamefont {Tomishige},\ and\ \citenamefont
  {Mizuno}}]{ariga2018nonequilibrium}%
  \BibitemOpen
  \bibfield  {author} {\bibinfo {author} {\bibfnamefont {T.}~\bibnamefont
  {Ariga}}, \bibinfo {author} {\bibfnamefont {M.}~\bibnamefont {Tomishige}},\
  and\ \bibinfo {author} {\bibfnamefont {D.}~\bibnamefont {Mizuno}},\
  }\bibfield  {title} {\bibinfo {title} {Nonequilibrium energetics of molecular
  motor kinesin},\ }\href@noop {} {\bibfield  {journal} {\bibinfo  {journal}
  {Physical review letters}\ }\textbf {\bibinfo {volume} {121}},\ \bibinfo
  {pages} {218101} (\bibinfo {year} {2018})}\BibitemShut {NoStop}%
\bibitem [{\citenamefont {Blickle}\ and\ \citenamefont
  {Bechinger}(2012)}]{blickle2012realization}%
  \BibitemOpen
  \bibfield  {author} {\bibinfo {author} {\bibfnamefont {V.}~\bibnamefont
  {Blickle}}\ and\ \bibinfo {author} {\bibfnamefont {C.}~\bibnamefont
  {Bechinger}},\ }\bibfield  {title} {\bibinfo {title} {Realization of a
  micrometre-sized stochastic heat engine},\ }\href@noop {} {\bibfield
  {journal} {\bibinfo  {journal} {Nature Physics}\ }\textbf {\bibinfo {volume}
  {8}},\ \bibinfo {pages} {143} (\bibinfo {year} {2012})}\BibitemShut {NoStop}%
\bibitem [{\citenamefont {Mart{\'\i}nez}\ \emph {et~al.}(2016)\citenamefont
  {Mart{\'\i}nez}, \citenamefont {Rold{\'a}n}, \citenamefont {Dinis},
  \citenamefont {Petrov}, \citenamefont {Parrondo},\ and\ \citenamefont
  {Rica}}]{martinez2016brownian}%
  \BibitemOpen
  \bibfield  {author} {\bibinfo {author} {\bibfnamefont {I.~A.}\ \bibnamefont
  {Mart{\'\i}nez}}, \bibinfo {author} {\bibfnamefont {{\'E}.}~\bibnamefont
  {Rold{\'a}n}}, \bibinfo {author} {\bibfnamefont {L.}~\bibnamefont {Dinis}},
  \bibinfo {author} {\bibfnamefont {D.}~\bibnamefont {Petrov}}, \bibinfo
  {author} {\bibfnamefont {J.~M.}\ \bibnamefont {Parrondo}},\ and\ \bibinfo
  {author} {\bibfnamefont {R.~A.}\ \bibnamefont {Rica}},\ }\bibfield  {title}
  {\bibinfo {title} {Brownian carnot engine},\ }\href@noop {} {\bibfield
  {journal} {\bibinfo  {journal} {Nature physics}\ }\textbf {\bibinfo {volume}
  {12}},\ \bibinfo {pages} {67} (\bibinfo {year} {2016})}\BibitemShut {NoStop}%
\bibitem [{\citenamefont {Filliger}\ and\ \citenamefont
  {Reimann}(2007)}]{filliger2007brownian}%
  \BibitemOpen
  \bibfield  {author} {\bibinfo {author} {\bibfnamefont {R.}~\bibnamefont
  {Filliger}}\ and\ \bibinfo {author} {\bibfnamefont {P.}~\bibnamefont
  {Reimann}},\ }\bibfield  {title} {\bibinfo {title} {Brownian gyrator: A
  minimal heat engine on the nanoscale},\ }\href@noop {} {\bibfield  {journal}
  {\bibinfo  {journal} {Physical Review Letters}\ }\textbf {\bibinfo {volume}
  {99}},\ \bibinfo {pages} {230602} (\bibinfo {year} {2007})}\BibitemShut
  {NoStop}%
\bibitem [{\citenamefont {Chiang}\ \emph {et~al.}(2017)\citenamefont {Chiang},
  \citenamefont {Lee}, \citenamefont {Lai},\ and\ \citenamefont
  {Chen}}]{chiang2017electrical}%
  \BibitemOpen
  \bibfield  {author} {\bibinfo {author} {\bibfnamefont {K.-H.}\ \bibnamefont
  {Chiang}}, \bibinfo {author} {\bibfnamefont {C.-L.}\ \bibnamefont {Lee}},
  \bibinfo {author} {\bibfnamefont {P.-Y.}\ \bibnamefont {Lai}},\ and\ \bibinfo
  {author} {\bibfnamefont {Y.-F.}\ \bibnamefont {Chen}},\ }\bibfield  {title}
  {\bibinfo {title} {Electrical autonomous brownian gyrator},\ }\href@noop {}
  {\bibfield  {journal} {\bibinfo  {journal} {Physical Review E}\ }\textbf
  {\bibinfo {volume} {96}},\ \bibinfo {pages} {032123} (\bibinfo {year}
  {2017})}\BibitemShut {NoStop}%
\bibitem [{\citenamefont {Chang}\ \emph {et~al.}(2021)\citenamefont {Chang},
  \citenamefont {Lee}, \citenamefont {Lai},\ and\ \citenamefont
  {Chen}}]{chang2021autonomous}%
  \BibitemOpen
  \bibfield  {author} {\bibinfo {author} {\bibfnamefont {H.}~\bibnamefont
  {Chang}}, \bibinfo {author} {\bibfnamefont {C.-L.}\ \bibnamefont {Lee}},
  \bibinfo {author} {\bibfnamefont {P.-Y.}\ \bibnamefont {Lai}},\ and\ \bibinfo
  {author} {\bibfnamefont {Y.-F.}\ \bibnamefont {Chen}},\ }\bibfield  {title}
  {\bibinfo {title} {Autonomous brownian gyrators: A study on gyrating
  characteristics},\ }\href@noop {} {\bibfield  {journal} {\bibinfo  {journal}
  {Physical Review E}\ }\textbf {\bibinfo {volume} {103}},\ \bibinfo {pages}
  {022128} (\bibinfo {year} {2021})}\BibitemShut {NoStop}%
\bibitem [{\citenamefont {Argun}\ \emph {et~al.}(2017)\citenamefont {Argun},
  \citenamefont {Soni}, \citenamefont {Dabelow}, \citenamefont {Bo},
  \citenamefont {Pesce}, \citenamefont {Eichhorn},\ and\ \citenamefont
  {Volpe}}]{argun2017experimental}%
  \BibitemOpen
  \bibfield  {author} {\bibinfo {author} {\bibfnamefont {A.}~\bibnamefont
  {Argun}}, \bibinfo {author} {\bibfnamefont {J.}~\bibnamefont {Soni}},
  \bibinfo {author} {\bibfnamefont {L.}~\bibnamefont {Dabelow}}, \bibinfo
  {author} {\bibfnamefont {S.}~\bibnamefont {Bo}}, \bibinfo {author}
  {\bibfnamefont {G.}~\bibnamefont {Pesce}}, \bibinfo {author} {\bibfnamefont
  {R.}~\bibnamefont {Eichhorn}},\ and\ \bibinfo {author} {\bibfnamefont
  {G.}~\bibnamefont {Volpe}},\ }\bibfield  {title} {\bibinfo {title}
  {Experimental realization of a minimal microscopic heat engine},\ }\href@noop
  {} {\bibfield  {journal} {\bibinfo  {journal} {Physical Review E}\ }\textbf
  {\bibinfo {volume} {96}},\ \bibinfo {pages} {052106} (\bibinfo {year}
  {2017})}\BibitemShut {NoStop}%
\bibitem [{\citenamefont {Marcinkiewicz}(1939)}]{marcinkiewicz1939propriete}%
  \BibitemOpen
  \bibfield  {author} {\bibinfo {author} {\bibfnamefont {J.}~\bibnamefont
  {Marcinkiewicz}},\ }\bibfield  {title} {\bibinfo {title} {Sur une
  propri{\'e}t{\'e} de la loi de gauss},\ }\href@noop {} {\bibfield  {journal}
  {\bibinfo  {journal} {Mathematische Zeitschrift}\ }\textbf {\bibinfo {volume}
  {44}},\ \bibinfo {pages} {612} (\bibinfo {year} {1939})}\BibitemShut
  {NoStop}%
\bibitem [{\citenamefont {Li}\ \emph {et~al.}(2019)\citenamefont {Li},
  \citenamefont {Horowitz}, \citenamefont {Gingrich},\ and\ \citenamefont
  {Fakhri}}]{li2019quantifying}%
  \BibitemOpen
  \bibfield  {author} {\bibinfo {author} {\bibfnamefont {J.}~\bibnamefont
  {Li}}, \bibinfo {author} {\bibfnamefont {J.~M.}\ \bibnamefont {Horowitz}},
  \bibinfo {author} {\bibfnamefont {T.~R.}\ \bibnamefont {Gingrich}},\ and\
  \bibinfo {author} {\bibfnamefont {N.}~\bibnamefont {Fakhri}},\ }\bibfield
  {title} {\bibinfo {title} {Quantifying dissipation using fluctuating
  currents},\ }\href@noop {} {\bibfield  {journal} {\bibinfo  {journal} {Nature
  communications}\ }\textbf {\bibinfo {volume} {10}},\ \bibinfo {pages} {1}
  (\bibinfo {year} {2019})}\BibitemShut {NoStop}%
\bibitem [{\citenamefont {Chaichian}\ and\ \citenamefont
  {Demichev}(2018)}]{chaichian2018path}%
  \BibitemOpen
  \bibfield  {author} {\bibinfo {author} {\bibfnamefont {M.}~\bibnamefont
  {Chaichian}}\ and\ \bibinfo {author} {\bibfnamefont {A.}~\bibnamefont
  {Demichev}},\ }\href@noop {} {\emph {\bibinfo {title} {Path integrals in
  physics: Volume I stochastic processes and quantum mechanics}}}\ (\bibinfo
  {publisher} {CRC Press},\ \bibinfo {year} {2018})\BibitemShut {NoStop}%
\bibitem [{\citenamefont {Kirsten}\ and\ \citenamefont
  {McKane}(2003)}]{kirsten2003functional}%
  \BibitemOpen
  \bibfield  {author} {\bibinfo {author} {\bibfnamefont {K.}~\bibnamefont
  {Kirsten}}\ and\ \bibinfo {author} {\bibfnamefont {A.~J.}\ \bibnamefont
  {McKane}},\ }\bibfield  {title} {\bibinfo {title} {Functional determinants by
  contour integration methods},\ }\href@noop {} {\bibfield  {journal} {\bibinfo
   {journal} {Annals of Physics}\ }\textbf {\bibinfo {volume} {308}},\ \bibinfo
  {pages} {502} (\bibinfo {year} {2003})}\BibitemShut {NoStop}%
\bibitem [{Note6()}]{Note6}%
  \BibitemOpen
  \bibinfo {note} {The entropy currents of the gyrator with harmonic potential,
  where the dynamical equations are linear, can be identified using standard
  approaches \cite
  {argun2017experimental,li2019quantifying,manikandan2021quantitative}. However
  the entropy currents for the anharmonic gyrators is non-trivial to find
  analytically due to indelible non-linearities present in the governing
  dynamical equation. To overcome this difficulty we use the short-time
  inference scheme \cite {das2022inferring}.}\BibitemShut {Stop}%
\bibitem [{Mat()}]{MathematicaMarkov}%
  \BibitemOpen
  \href@noop {} {\bibinfo {title} {Wolfram research (2012),
  continuousmarkovprocess, wolfram language function, link =
  {\url{https://reference.wolfram.com/language/ref/continuousmarkovprocess.html}}}}\BibitemShut
  {NoStop}%
\end{thebibliography}
\providecommand{\noopsort}[1]{}\providecommand{\singleletter}[1]{#1}%

\beginsupplement

\title{Supplementary Information: Non-monotonic skewness of currents in non-equilibrium steady states}
\author{Sreekanth K Manikandan$^\dagger$}\email{sreekanth.km@fysik.su.se}
\affiliation{NORDITA, KTH Royal institute of technology and Stockholm university, Stockholm.}
\author{Biswajit Das}\thanks{These authors contributed equally}
\author{Avijit Kundu}\thanks{These authors contributed equally}
\author{Raunak Dey}
\affiliation{Department of Physical Sciences, IISER Kolkata}
\author{Ayan Banerjee}\email{ayan@iiserkol.ac.in}
\affiliation{Department of Physical Sciences, IISER Kolkata}%
\author{Supriya Krishnamurthy}\email{supriya@fysik.su.se}
\affiliation{Department of Physics, Stockholm university, Stockholm.}

\maketitle
\onecolumngrid
\section*{Supplemental material}
In this section, we reproduce the exact calculation of the moment generating function of $\Delta S_{tot} (t)$ for the Stochastic Sliding Parabola model, previously obtained in \cite{manikandan2018exact}.  

\section{Exact calculation of the MGF of $\Delta S_{tot}(t)$ for the Stochastic Sliding Parabola (SSP) model}
\label{slip:ss}
The stationary probability distribution for $x$ and $\lambda$  is given by \cite{Pal:2013wfb}, 
\begin{align}
\label{slidp:pst}
p_{st}(x(t),\;\lambda(t))&=\frac{\exp \left(-\frac{(\delta+1) \left(\delta^2\theta (x-\lambda)^2+\delta \left(\theta x^2+\lambda^2\right)+\lambda^2\right)}{2D\tau_0\theta \left(\delta^2 (\theta+1)+2 \delta+1\right)}\right)}{2 \pi  \sqrt{\frac{D^2\tau_0^2\theta \left(\delta^2 (\theta+1)+2 \delta+1\right)}{\delta (\delta+1)^2}}}.
\end{align}
the MGF of total entropy production $\Delta S_{tot}(t)$ can be written down in the following manner. First, the joint probability density functional of trajectories starting at $t=0$ at $(x_0,\lambda_0)$ and ending at $t=\tau$ at $(x_t,\lambda_t)$ may be written as,
\begin{equation}
\label{slidp:pxl}
P[x(\cdot),\lambda(\cdot)]={\bm N}\; \rm{exp}\; \bigg\lbrace-\int_{0}^{t}ds\;L(\dot{x}(s),x(s),\dot{\lambda}(s),\lambda(s),s)\bigg\rbrace
\end{equation}
with the Lagrangian,
\begin{equation}
\label{slidp:lagr}
L=\frac{1}{4D}\left(\;[\dot{x}+\frac{\delta(x-\lambda)}{\tau_0}]^2+\frac{1}{\theta}\;[\dot{\lambda}+\frac{\lambda}{\tau_0}]^2\right).
\end{equation}
The normalization constant {$\bm N$ } for this case is \cite{chaichian2018path},
\begin{equation}
\label{slidp:norm}
{\bm N}=\exp\left(\;\frac{1}{2}\;\left[\frac{\delta+1}{\tau_0}\right] t\;\right).
\end{equation}
The entropy production in the steady-state in the time interval $[0,\tau]$ for the SSP is then,
\begin{equation}
\label{slidp:ep}
\Delta S_{tot}(t)=\frac{\delta}{D\tau_0}\int_0^t ds\; \lambda(s)\;\dot{x}(s)+\frac{\delta^2 \left(\delta \left(\theta \left(x_0^2-x_t^2\right)+2 x_0 \lambda_0-2 x_t \lambda_t-\lambda_0^2+\lambda_t^2\right)+2 x_0 \lambda_0-2 x_t \lambda_t\right)}{2 D \tau_0 \left(\delta^2 (\theta+1)+2 \delta+1\right)}.
\end{equation}
This form of the entropy production can easily be obtained by equating it to the ratio of the probabilities of forward and time-reversed trajectories using Eq. (\ref{slidp:pst}) and Eq. (\ref{slidp:pxl}) and the form of the Lagrangian Eq. (\ref{slidp:lagr}). 
Hence, upto a normalization factor \textbf{C} (determined by Eq.\ \eqref{slidp:pst} and \eqref{slidp:norm}), we have the following expression for the MGF of $\Delta S_{tot}(t)$,
\begin{equation}
\label{slidp:pwd}
\langle e^{-\frac{u}{2}\;\Delta S_{tot}(t)}\rangle=\textbf{C} \int dx_0 \;\int d\lambda_0\;\int dx_t\;\int d\lambda_t\int_{x_0,\lambda_0}^{x_t,\lambda_t}\; \mathcal{D}[x(\cdot),\; \lambda(\cdot)]\;e^{-\beta \; S[\;x(\cdot),\; \lambda(\cdot),\;u\;]},\;
\end{equation}
with the augmented action
\begin{equation}
\label{slidp:Sss}
\begin{split}
S[\;x(\cdot),\; \lambda(\cdot),\;u\;]&= \frac{(\delta+1) \left(\delta^2\theta (x_0-\lambda_0)^2+\delta \left(\theta x_0^2+\lambda_0^2\right)+\lambda_0^2\right)}{2D\tau_0\theta \left(\delta^2 (\theta+1)+2 \delta+1\right)}\\ &+ \int_{0}^{t}ds\;\frac{1}{4D}\left(\;[\dot{x}+\frac{\delta(x-\lambda)}{\tau_0}]^2+\frac{1}{\theta}\;[\dot{\lambda}+\frac{\lambda}{\tau_0}]^2\right)+\frac{u}{2}\;\Delta S_{tot}(t)[x,\lambda].
\end{split}
\end{equation}
after several partial integrations, it can be shown that the above quadratic action reduces to
\begin{equation}
\label{bcs}
S[\;x(\cdot),\; \lambda(\cdot),\;u\;]=\frac{1}{4D}\;\left[\begin{array}{cc}x&\lambda
\end{array}\right]\;\textbf{A}_u\;\left[\begin{array}{c}x\\\lambda
\end{array}\right]+\text{Boundary terms in } (x, \lambda, u),
\end{equation}
where the kernel is defined by the operator: 
\begin{align}
\label{slidp:ele}
\textbf{A}_u &= \left[\begin{array}{cc}-\frac{d^2}{d 
s^2}+\frac{\delta^2}{\tau_0^2} & k\;\frac{\delta}{\tau_0}\frac{d}{d s}-\frac{\delta^2}{\tau_0^2}\\-k\;\frac{\delta}{\tau_0}\frac{d}{d s}-\frac{\delta^2}{\tau_0^2} & -\frac{1}{\theta}\frac{d^2}{d 
s^2}+\frac{1}{\theta \tau_0^2}+\frac{\delta^2}{\tau_0^2} \end{array} \right];& k&\equiv 1-u.
\end{align} 
Carrying out the Gaussian integral, and requiring the boundary terms to vanish, the generating function at arbitrary times $\tau$ can be written down as a ratio of functional determinants,
\begin{equation}
\label{slidp:detr1}
\langle e^{-\frac{u}{2}\; \Delta S_{tot}(t)} \rangle = \sqrt{\frac{\det \textbf{A}_{u=0}}{\det \textbf{A}_{u}}}\equiv \Phi(u).
\end{equation}
This ratio can be computed using a technique described in \cite{kirsten2003functional} and used in \cite{Manikandan:2017awd}, which is based on the spectral -$\zeta$ functions of Sturm-Liouville type operators. Applying this method, it can be shown that this ratio can be obtained in terms of a characteristic polynomial function $F$ as,
\begin{align}
\label{mgf}
\left\langle e^{-\frac{u}{2}\; \Delta S_{tot}(t)}\right\rangle &=\sqrt{\frac{F(1)}{F(k)}},  &F(k) &\equiv \text{Det}\left[M+N H(t)\right],
& k&=1-u,
\end{align}
where, $H$ is the matrix of suitably normalized fundamental solutions of the homogeneous equation, $\textbf{A}_u\; \vec{x}=0 $, and is defined as,
\begin{align}
H(t)&=
\left[
\begin{array}{cccc}
 x_1(t) & x_2(t) & x_3(t) & x_4(t) \\
 \lambda_1(t) & \lambda_2(t) & \lambda_3(t) & \lambda_4(t)\\
 \dot{x}_1(t) & \dot{x}_2(t) & \dot{x}_3(t) & \dot{x}_4(t)\\
\dot{\lambda}_1(t) & \dot{\lambda}_2(t) & \dot{\lambda}_3(t) & \dot{\lambda}_4(t) \\
\end{array}
\right],& H(0) &=\textbf{I}_4.
\end{align}
$M$ and $N$ have information about the boundary conditions from Eq.\ \eqref{bcs} and we require,
\begin{align}
M \left[\begin{array}{c}\vec{x}(0)\\\dot{\vec{x}}(0)\end{array}\right]&=0,&N \left[\begin{array}{c}\vec{x}(t)\\\dot{\vec{x}}(t)\end{array}\right]&=0.
\end{align}
A derivation of Eq.\ \eqref{mgf}, applicable to a class of driven Langevin systems with quadratic actions is given in \cite{Manikandan:2017awd}. We would also like to stress that, the expression given in Eq.\ \eqref{mgf} is valid only for  $u \in\left[u^-(\tau),u^+(\tau)\right] $ for which the operator $A_u$ doesn't have negative eigenvalues. The MGF is not analytic outside this interval.\par
For the SSP in the steady state, we find,
the four independent solutions of $A_u \vec{x}=0$ to be \begin{align}
\vec{x}_i&=\left[\begin{array}{c}x_i(t)\\\lambda_i(t)
\end{array}\right],&i&=1\text{ to }4,
\end{align} 
where,
\begin{align}
\lambda_i(t)&=\scalebox{1}{$\exp \left(\pm\;\frac{\tau  \sqrt{\frac{\delta ^2 \theta +\delta ^2+\delta ^2 \theta  \left(-(1-u)^2\right)\;\pm\;\tau_0^2 \sqrt{\frac{\delta ^4 \left(\theta -\theta  (1-u)^2+1\right)^2-2 \delta ^2 \left(\theta  \left((1-u)^2-1\right)+1\right)+1}{\tau_0^4}}+1}{\tau_0^2}}}{\sqrt{2}}\right)$},  \\
x_i(t)&=\scalebox{1}{$\frac{\tau_0 \left((u-1) \lambda_i '(t) \left(\delta ^2 \theta  (u-2) u-1\right)+\tau_0 \left(\delta  \lambda_i ''(t)+\tau_0 (u-1) \lambda_i ^{(3)}(t)\right)\right)+\delta  \lambda_i (t) \left(\delta ^2 \theta  (u-2) u-1\right)}{\delta ^3 \theta  (u-2) u}$}
\end{align}
Matrices $M$ and $N$ are given by,
\begin{align}
M&=\scalebox{1}{$\left(
\begin{array}{cccc}
 \frac{(1-(1-u) \theta ) \delta ^3+2 \delta ^2+\delta }{2 D \left((\theta +1) \delta ^2+2 \delta +1\right) \tau_0} & \frac{\delta  \left(((1-u) \theta -1) \delta ^2-u \delta -u+1\right)}{2 D \left((\theta +1) \delta ^2+2 \delta +1\right) \tau_0} & -\frac{1}{2 D} & 0 \\
 -\frac{(2-u) \delta ^2 (\delta +1)}{2 D \left((\theta +1) \delta ^2+2 \delta +1\right) \tau_0} & \frac{(2-u) \theta  \delta ^3+(\theta +1) \delta ^2+2 \delta +1}{2 D \theta  \left((\theta +1) \delta ^2+2 \delta +1\right) \tau_0} & 0 & -\frac{1}{2 D \theta } \\
 0 & 0 & 0 & 0 \\
 0 & 0 & 0 & 0 \\
\end{array}
\right)$}\\
N&=\scalebox{1}{$\left(
\begin{array}{cccc}
 0 & 0 & 0 & 0 \\
 0 & 0 & 0 & 0 \\
 \frac{\delta  \left(((1-u) \theta +1) \delta ^2+2 \delta +1\right)}{2 D \left((\theta +1) \delta ^2+2 \delta +1\right) \tau_0} & -\frac{\delta  \left((1-u) \theta  \delta ^2+\delta ^2+(1-u) \delta +\delta -u+1\right)}{2 D \left((\theta +1) \delta ^2+2 \delta +1\right) \tau_0} & \frac{1}{2 D} & 0 \\
 -\frac{u \delta ^2 (\delta +1)}{2 D \left((\theta +1) \delta ^2+2 \delta +1\right) \tau_0} & \frac{(\theta -(1-u) \theta ) \delta ^3+(\theta +1) \delta ^2+2 \delta +1}{2 D \theta  \left((\theta +1) \delta ^2+2 \delta +1\right) \tau_0} & 0 & \frac{1}{2 D \theta } \\
\end{array}
\right)$}
\end{align} \par
Using these, the MGF can be computed exactly using Eq.\ \eqref{mgf}, and various moments of the probability distribution can also be exactly obtained for any $t$. 
\subsection{Dependence of $\tau_E$ on system parameters}
\begin{figure}
    \centering
    \includegraphics[scale=0.3]{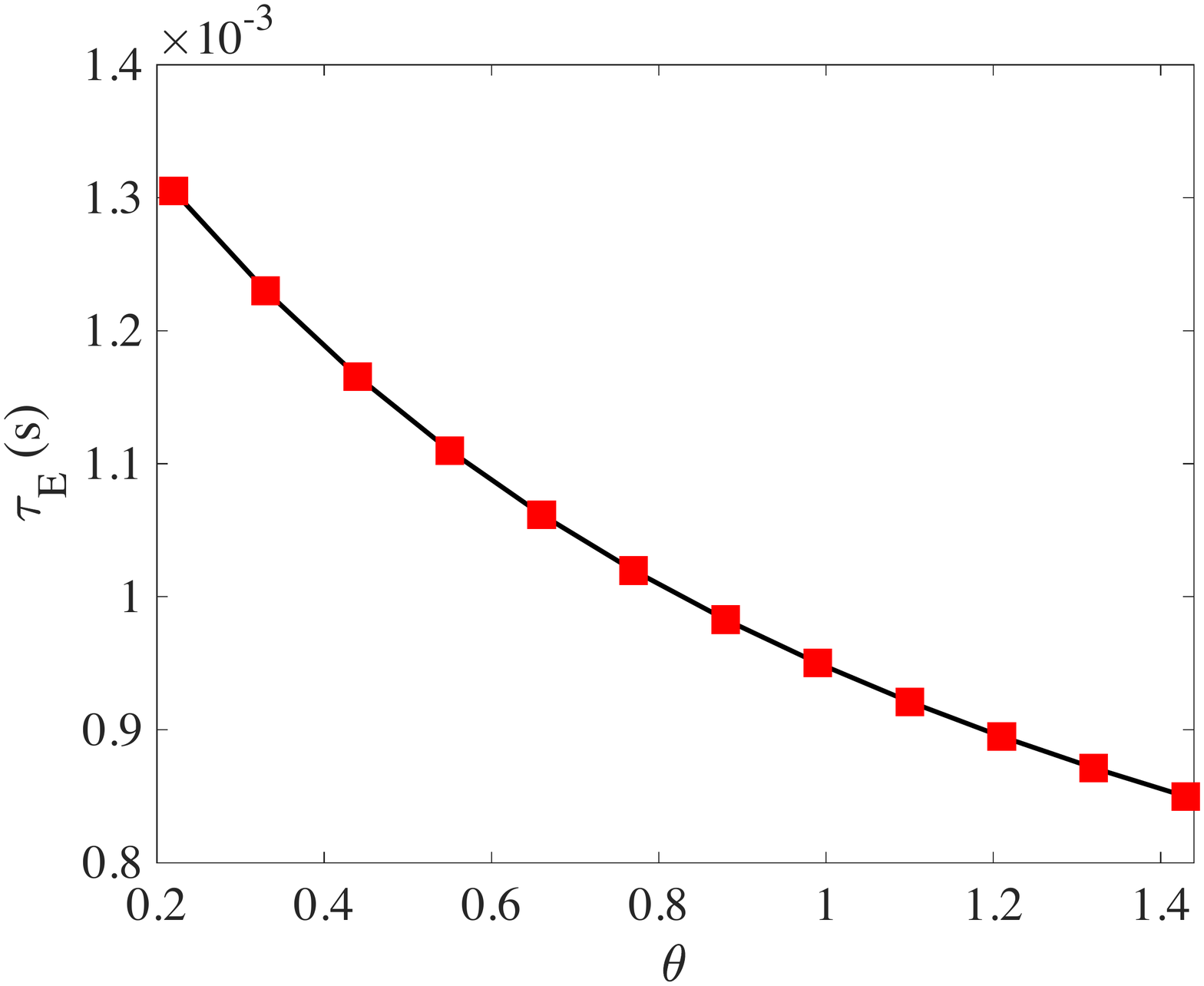}
    \includegraphics[scale=0.42]{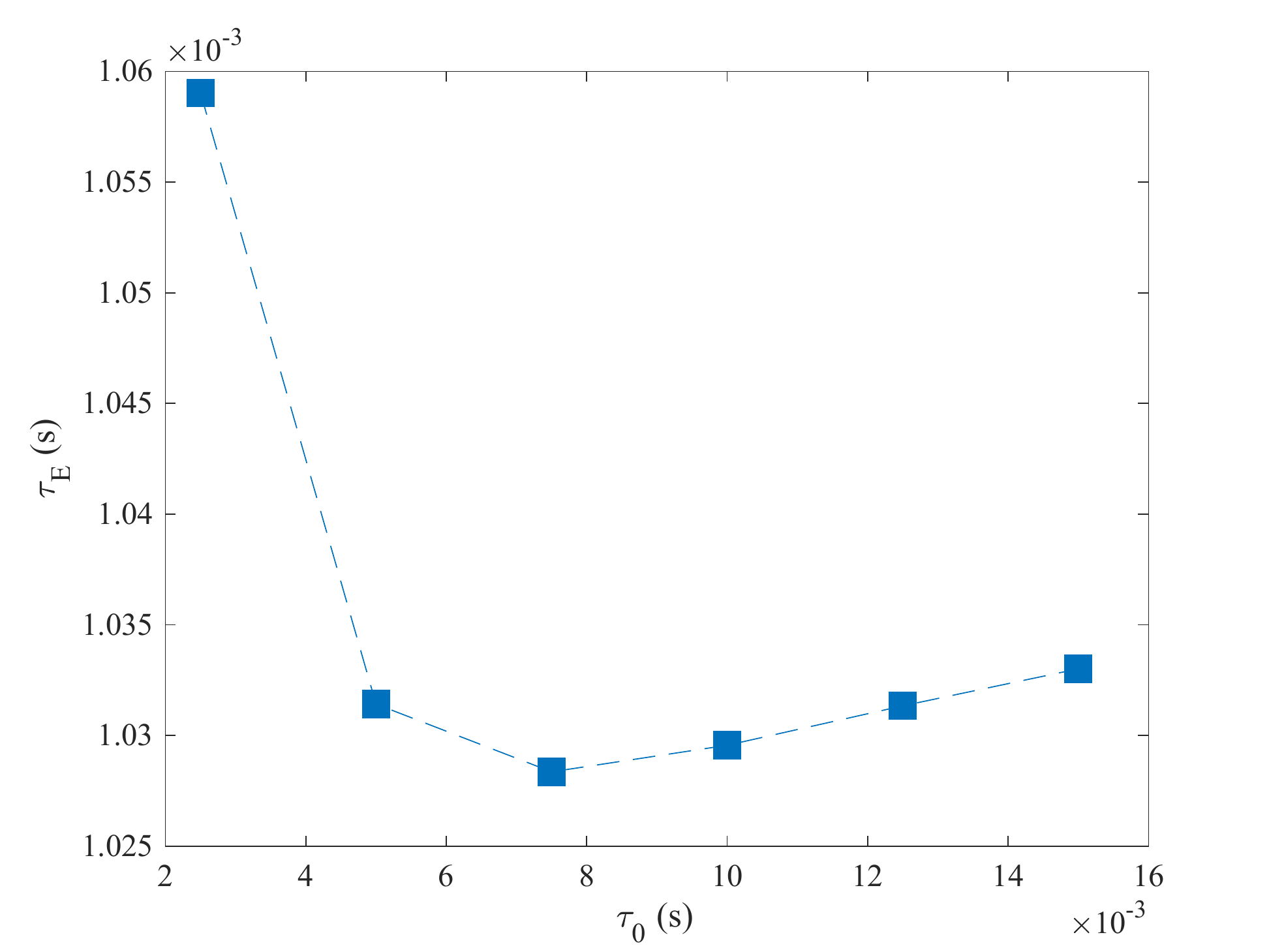}
    \caption{ Dependence of $\tau_E$ on the relative magnitude of the Ornstein Uhlenbeck driving $\theta$. We find that $\tau_E$ monotonically decreases with increasing $\theta$, which is the limit at which the system goes further away from equilibrium. b) Dependence of $t_E$ on $\tau_0$. We find that $\tau_E$ is a non-monotonic function of $\tau_0$ and features a minimum at a particular $\tau_0$ value.
    }
    \label{fig:3}
\end{figure}

For the stochastic sliding parabola model, it is possible to analytically compute how $\tau_E$, the time at which Skewness peaks, depends on the system parameters, using the exact expressions. In Fig.\ \ref{fig:3}, we demonstrate how $\tau_E$ depends on $\theta$ and $\tau_0$. We find that $\tau_E$ monotonically decreases with increasing $\theta$, which is the limit at which the system goes further away from equilibrium. This means, further away the system is from equilibrium, a higher resolution in time will be required to see the nonmonotonicity in current fluctuations. The dependence of $\tau_E$ on $\tau_0$ is found to be more non-trivial. We find that $\tau_E$ is a non-monotonic function of $\tau_0$ and features a minimum at a particular $\tau_0$ value.

\section{Entropy currents of anharmonic Brownian gyrators}
\label{brgr:ss}
The brownian gyrator is one of the minimal prototypes of a microscopic heat engine \cite{filliger2007brownian,argun2017experimental,chiang2017electrical}. It consists of a micron-sized particle trapped in a generic potential well, coupled to two heat reservoirs with different temperatures along two orthogonal directions. When the two degrees of freedom of the trapped particle are coupled, it can be shown that the system reaches a non-equilibrium stationary state, where the particle starts gyrating around the minima of the potential. The dynamics of the system, in the overdamped limit, can be expressed in terms of coupled \textit{langevin} equations: 

\begin{equation}
    \gamma \dot{x} = -\frac{\partial U(x,y)}{\partial x} + \sqrt{2 \gamma k_B T_1} \eta_1(t)
    \label{eq:lan1}
\end{equation}
\begin{equation}
    \gamma \dot{y} = -\frac{\partial U(x,y)}{\partial y} + \sqrt{2 \gamma k_B T_2} \eta_2(t)
    \label{eq:lan2}
\end{equation}

Here, $U(x,y)$ denotes the confining potential in the $x-y$ plane. The $x$ axis is coupled to a thermal reservoir at temperature $T_1$ and the $y$ axis is coupled to another thermal reservoir at temperature $T_2$. The corresponding thermal noises $\eta_i(t)$ are of Gaussian nature and \textit{white} in time, such that $\langle \eta_i(t)\rangle = 0$ and $\langle \eta_i(t)\eta_j(t^\prime)\rangle = \delta_{ij}\delta(t-t^\prime)$. The viscous drag of the medium is denoted by $\gamma$, which is related to the temperatures of the reservoirs through the Einstein relation, $D_i \gamma = k_B T_i$, $k_B$ is the Boltzmann constant (We set $k_B$ = 1 for simplicity). In this work, we choose the anharmonic Brownian gyrator, where the confining potential is anharmonic \cite{chang2021autonomous,das2022inferring}, as an example of a non-linear diffusive model.\\
 
We first consider a Brownian gyrator with a double-well confining potential of the form,
\begin{equation}
    U_{dw}(x',y') = x'^4 - 2bx'^2 + \frac{1}{2}ky'^2
    \label{eq:dw}
\end{equation}

Where axes of the potential $x'$ and $y'$ are rotated by an angle $\theta$ with respect to the temperature axes $(x,y)$ [ axes of the co-ordinate frame] as,
\begin{equation}
    \begin{bmatrix}
    x'\\y'
    \end{bmatrix}
    = \begin{bmatrix}
    \cos \theta  & -\sin \theta \\
    \sin \theta  &   \cos \theta
    \end{bmatrix}
    \times 
    \begin{bmatrix}
    x \\y
    \end{bmatrix}
    \label{eq:rot}
\end{equation}

The parameter `$b$' can be used to tune the bi-stable nature of the potential along the $x'$ direction as the barrier height ($=b^2$)  and the position of the minima (($= \pm \sqrt{b}$) ) of the potential are dependent on it. The stiffness constant `$k$' characterises the harmonic part of the potential along the $y'$ direction. We choose $b = 1$, $k = 2$, and $\theta = 45^\circ$ for the analysis performed in this work. \\

\begin{figure*}[h]
    \centering
    \includegraphics[width=\textwidth]{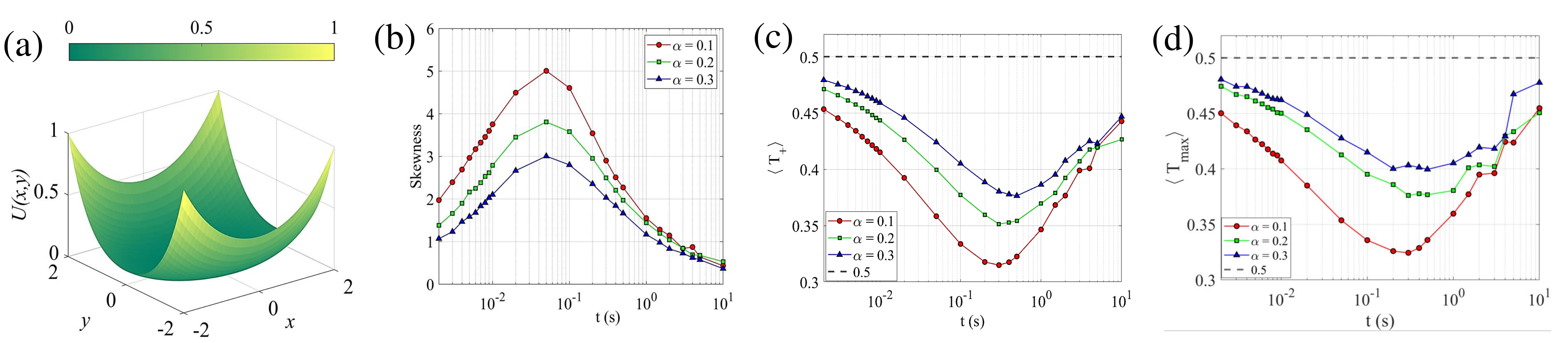}
    \caption{a) Quartic confining potential with `$k_1$' = 1 \ \& `$k_2$'  =2. (b) Skewness ,(c) $\langle T_+ \rangle$ and d) $\langle T_{max} \rangle$ of the entropy currents as a function of $t$, obtained numerically for the gyrator model with a quartic confining potential for different values of the ratio of the temperatures ($\alpha = \frac{T_2}{T_1}$).}
    \label{fig:plot_brgr_qu}
\end{figure*}

We also consider another anharmonic Brownian gyrator with a quartic confining potential given by,
\begin{equation}
    U_{qw}(x',y') = (k_1 x'^2 + k_2 y'^2)^2
\label{eq:qw}
\end{equation}
The parameters ~`$k_1$' and `$k_2$' can be considered as the stiffness constants along the respective directions. We set $k_1 = 1$, $k_2 = 2$ along with $\theta = 45^\circ$ to construct an anisotropic quartic potential for the analysis performed in this work. \\

The entropy currents for both gyrators are constructed numerically with third-order polynomial basis functions using the TUR-based short-time inference scheme \cite{manikandan2020inferring,shun,tan}, as described in Ref.\cite{das2022inferring} \footnote{The entropy currents of the gyrator with harmonic potential, where the dynamical equations are linear, can be identified using standard approaches \cite{argun2017experimental,li2019quantifying,manikandan2021quantitative}. However the entropy currents for the anharmonic gyrators is non-trivial to find analytically due to indelible non-linearities present in the governing dynamical equation. To overcome this difficulty we use the short-time inference scheme \cite{das2022inferring}.}. We have looked into the properties of entropy currents corresponding to the different non-equilibrium conditions controlled by the ratios of the temperatures ($\alpha = T_2/T_1$) of the thermal reservoirs of the systems, where $\alpha = 0.1$ corresponds to the most non-equilibrium configuration we considered, and $\alpha = 0.3$ is the least non-equilibrium configuration of both systems. The non-monotonic features of the entropy current fluctuations for the gyrator system with double-well confining potential are  shown in Fig. \ref{fig:plot_nonlinear} of the main text and in Fig. \ref{fig:plot_brgr_qu} we show the similar results for the quartic well confining potential.

\section{Skewness of currents in continuous time Markov jump processes}
\label{mjump:ss}

Here we demonstrate that Skewness $\propto t^{-1/2}$ in the $t\rightarrow 0$ limit for discrete space system evolving according to a continuous time Markov jump process. We consider a set of $M$ number of states  $ \lbrace i \rbrace$, $i=1,\; 2,\; ... M$,  and transition rates $\Gamma_{ij}\geq$. Let $\pi(i)$ correspond to the steady state probability of finding the system in the state $i$. A stochastic realization of the system is denoted by $x(s)\in \lbrace i \rbrace$ with $s\in \left[ 0,t\right)$. The fluctuating current between any two pairs of states $i$ and $j$ can be computed as,
\begin{align}
    j_{ij}(t)=\sum_k \delta (s-s_k)\left(\delta_{x(s_k^+),j}\delta_{x(s_k^-),i}-\delta_{x(s_k^+),i}\delta_{x(s_k^-),j} \right)
\end{align}
where $x(s_k^+) \; (x(s_k^-))$ corresponds to the state of the system immediately after (before) the transition at times $s=s_k$. A generalized time-integrated current in this system is given by, 
\begin{align}
    J(t)=\int_{0}^t ds \; \sum_{i<j} d_{ij}\; j_{ij}(s)
\end{align},
where $d_{ij}=-d_{ji}$ are weighting factors which are constants. The steady state average of this current is given by, 
\begin{align}
    \langle J (t) \rangle = t\sum_{ij}\pi(i)\Gamma_{ij}d_{ij}
\end{align}
A particular choice, $d_{ij}=F_{ij} =\log \frac{\Gamma_{ij}\pi(i)}{\Gamma_{ji}\pi(j)}$ corresponds to the current $J=\Delta S_{tot}(t)$. 

The fluctuations of any such current $J$ can be calculated using it's moment generating function, 
\begin{align}
    G(\lambda,t) = \langle e^{-\lambda J (t)} \rangle = \langle 1 \vert e^{t \mathcal{L}(\lambda)}\vert \pi\rangle,
\end{align}
where $\mathcal{L}(\lambda)$ is the tilted transition matrix with elements,
\begin{align}
    \mathcal{L}_{ij}(\lambda) = \Gamma_{ji}\exp(\lambda d_{ji})-\delta_{ij}\sum_l \Gamma_{il}.
\end{align}
We are particularly interested in the small time properties of the moments of the current $J(t)$. This can be obtained by Taylor expanding $G$ as near $t=0$. Keeping to first order in $t$, we obtain, 
\begin{align}
    G(\lambda,t)\sim = 1 - t\; \langle 1 \vert \mathcal{L}_{ij}(\lambda) \vert \pi \rangle+ \mathcal{O}[t^2]
\end{align}
For a generic choice of $\Gamma_{ij}$ and $d_{ij}$, it is possible to verify that all the moments of $J$, obtained by Taylor expanding $G$ as a function of $\lambda$, will be proportional to $t$ for small $t$.
Thus the skewness as defined in the maintext, will be proportional to $t^{-\frac{1}{2}}$ for $t \rightarrow 0$. 

In Fig.\ (4)a of the maintext, we show this for a three-level system $x\in \lbrace 0,1,2\rbrace$, with energy levels $E_i =\lbrace 0,E_1,E_2 \rbrace$. We have assumed that the transitions between the levels $0$ and $1$, and the levels $0$ and $2$ are mediated by a hot reservoir at inverse temperature $\beta_1 = \frac{1}{k_B T_1}$, where $k_B$ is the Boltzmann constant and $T_1$ is the temperature of the hot  reservoir. The corresponding transition rates obey the local detailed balance condition:
\begin{align}
    \Gamma_{01}&=\exp(-\beta_1 E_1) \Gamma_{10},& \Gamma_{02}&=\exp(-\beta_1 E_2) \Gamma_{20}.
\end{align}
The transitions between the levels $1$ and $2$ are assumed to be mediated by a cold reservoir at inverse temperature $\beta_2 = \frac{1}{k_B T_2}>\beta_1$. The corresponding transition rates obey
\begin{align}
    \Gamma_{12}&=\exp(-\beta_2 (E_2-E_1)) \Gamma_{21}.
\end{align}
The skewness of arbitrary currents, and in particular entropy production can be straightforwardly computed for this model using the expressions given above. There are also standard techniques available to numerically simulate this model as a continuous time Markov process \cite{MathematicaMarkov}, and to obtain the statistics of fluctuating currents as a function of time. The plots in Fig.\ 4 of main text are obtained for the parameter choices $E_1=k_B T_1$, $E_2=2 k_BT_1$, $\Gamma_{01}=\Gamma_{02}=\Gamma_{12} = 1s^{-1}$, $\beta_1 =(k_B T_1)^{-1}$ and $T_2 = T_1/2, \;T_1/3 \text{ and } T_1/5$, for which the entropy production rates are, $\sigma = 0.027 k_B s^{-1}$, $0.076 k_B s^{-1}$ and $0.176 k_B s^{-1}$ respectively. 

\section{Parameter Values}
\label{param:ss}
\begin{itemize}
\item{Figure 1:} $\tau \equiv \frac{1}{2\pi f_c} = 0.0013~s$, $\tau_0 = 0.0025~s$, $D = 1.6452 \times 10^{-13}~m^2/s, A = 0.3 \times (0.6 \times 10^{-6})^2 ~m^2/s$.

\item{Figure 2:} (a) - (c): $\tau \equiv \frac{1}{2\pi f_c} = 0.0013~s$, $\tau_0 = 0.0025~s$, $D = 1.6452 \times 10^{-13}~m^2/s $,  $A = [0.1,\;0.15,\;0.2,\;0.25,\;0.3,\;0.35]\times (0.6\times 10^{-6})^2~m^2/s$;  
(b) - (d): $\tau \equiv \frac{1}{2\pi f_c} = 0.0013~s$, $\tau_0 = 0.0025~s$, $D = 1.6452 \times 10^{-13}~m^2/s $,  $A = [0.1,\;0.2,\;0.3]\times (0.6\times 10^{-6})^2~m^2/s$.

\item{Figure 3:} (a): $b = 1, k = 2$; (b) - (d): $\gamma = 1$, $T_1 \equiv D_1 = 1$, $\alpha \equiv \frac{T_2}{T_1} = [0.1, \;0.2, \;0.3]$.

\item{Figure 4:} $E_1=k_B T_1$, $E_2=2 k_BT_1$, $\Gamma_{01}=\Gamma_{02}=\Gamma_{12} = 1s^{-1}$, $\beta_1 =(k_B T_1)^{-1}$, $T_2 = [\frac{1}{2},\; \frac{1}{3},\;\frac{1}{5}] \times T_1 $.

\item {Figure S2:} (a): $k_1 = 1, k_2 = 2$; (b) - (d): $\gamma = 1$, $T_1 \equiv D_1 = 1$, $\alpha \equiv \frac{T_2}{T_1} = [0.1, \;0.2, \;0.3]$.

\end{itemize}

\end{document}